\documentclass[useAMS,usenatbib]{mn2e}
\usepackage[pdftex]{graphicx}
\usepackage{amsmath}
\usepackage{morefloats}
\usepackage[varg]{txfonts}
\def\kms{km\,s$^{-1}$}

\title[Water and methanol in low-mass protostellar outflows: gas-phase synthesis, ice sputtering and destruction]{Water and methanol in low-mass protostellar outflows: gas-phase synthesis, ice sputtering and destruction}
\author[A.N. Suutarinen et al.]{
A.N. Suutarinen$^{1}$\thanks{E-mail:aleksi.suutarinen@iki.fi},
L.E. Kristensen$^{2}$,
J.C. Mottram$^{3}$,
H.J. Fraser$^{1}$ and\newauthor
E.F. van Dishoeck$^{3,4}$\\
$^{1}$The Open University, Department of Physical Sciences, Milton Keynes, MK7 6AA, United Kingdom\\
$^{2}$Harvard-Smithsonian Center for Astrophysics, 60 Garden Street, Cambridge, MA 02138, USA \\
$^{3}$Leiden Observatory, Leiden University, PO Box 9513, 2300 RA Leiden, The Netherlands\\
$^{4}$Max-Planck-Institut f\"{u}r Extraterrestrische Physik, Giessenbachstrasse 1, 85748 Garching, Germany\\
}

\begin{document}

\date{Accepted 2014 February 27. Received 2014 February 27; in original form 2013 December 19.}

\pagerange{\pageref{firstpage}--\pageref{lastpage}} \pubyear{2014}

\maketitle

\label{firstpage}

\begin{abstract}
Water in outflows from protostars originates either as a result of gas-phase synthesis from atomic oxygen at $T$ $\ga$ 200 K, or from sputtered ice mantles containing water ice. We aim to quantify the contribution of the two mechanisms that lead to water in outflows, by comparing observations of gas-phase water to methanol (a grain surface product) towards three low-mass protostars in NGC1333. In doing so, we also quantify the amount of methanol destroyed in outflows. To do this, we make use of \textit{JCMT} and \textit{Herschel}-HIFI data of H$_2$O, CH$_3$OH and CO emission lines and compare them to \textsc{RADEX} non-LTE excitation simulations. We find up to one order of magnitude decrease in the column density ratio of CH$_3$OH over H$_2$O as the velocity increases in the line wings up to $\sim$15 \kms. An independent decrease in $X({\rm CH_3OH})$ with respect to CO of up to one order of magnitude is also found in these objects. We conclude that gas-phase formation of H$_2$O must be active at high velocities (above 10 \kms\, relative to the source velocity) to re-form the water destroyed during sputtering. In addition, the transition from sputtered water at low velocities to formed water at high velocities must be gradual. We place an upper limit of two orders of magnitude on the destruction of methanol by sputtering effects.
\end{abstract}


\section{Introduction}
\label{sec:intro}
Water plays a unique role in probing the physical and chemical conditions of star-forming regions. In the cold regions of protostars, water is frozen out as ice covering the dust grains, with only trace amounts found in the gas phase \citep[$\la$10$^{-7}$;][Schmalzl et al. in prep.]{Caselli2012,Herpin2012,Mottram2013}. The molecular outflows launched by the accreting protostar form an interesting laboratory for studying water in protostellar systems: the high water abundances observed in outflows \citep[$>$ 10$^{-6}$;][]{Nisini2010,Vasta2012, Santangelo2012, Bjerkeli2012, Dionatos2013,VanLoo2013} arise from a combination of sputtering of ice mantles and direct gas-phase synthesis, however the relative water abundance contributions are unknown. Constraining the relative contributions of these two routes is important for using water as a diagnostic of the physical conditions in outflows, as well as constraining the chemistry in the various parts of the outflow.

Water formation on dust grains is a complicated process
\citep[e.g.][]{Ioppolo2008,Lamberts2013}, but generally
involves the repeated hydrogenation of oxygen-bearing
species such as O, OH and O$_2$ \citep{Tielens1982}.
The only efficient gas-phase formation route for H$_2$O is a
high-temperature ($T \ga 200$ K) neutral-neutral
reaction chain \citep[e.g.][]{Draine1983, Bergin1999, Tielens2005,
Glassgold2009}:
\begin{equation}
{\rm O \xrightarrow{H_2} OH \xrightarrow{H_2} H_2O + H}.
\end{equation}

Another common grain-surface material, methanol, can be
formed on dust grains through through
repeated hydrogenation of CO \citep[e.g.][]{Tielens1982,Tielens2005}
\begin{equation}
{\rm CO_{surface} \xrightarrow{H} HCO_{surface} \xrightarrow{H}
CH_3O_{surface} \xrightarrow{H} CH_3OH_{surface}}.
\end{equation}
Unlike water, gas-phase reactions that produce CH$_3$OH are very slow compared to grain
surface chemistry for any physical conditions expected within protostellar systems \citep{Geppert2006}.

Both water \citep[e.g.][]{Codella2010,vanDishoeck2011}
and methanol \citep[e.g.][]{Bachiller1995,Kristensen2010}
have been detected in the gas phase of
the outflows and shocks of young stellar objects (YSOs).
In astrophysical environments the grain surface
reaction chain leads to
grain surface abundances of order $10^{-6}-10^{-5}$ for
CH$_3$OH with respect to gas-phase H$_2$ \citep[e.g.][]{VanderTak2000, Cuppen2009}. This results in
relative abundances of CH$_3$OH ice with respect to H$_2$O ice ranging
from a few to $\sim$ 30 \%
\citep[e.g.][]
{Dartois1999,Pontoppidan2003,Pontoppidan2004,Gibb2004,Boogert2008,Oberg2011}.
Models of cold dense clouds by \citet{LeeH.-H.1996}, which only include gas-phase production, result in
gas-phase abundances of CH$_3$OH no higher than between
$10^{-10}$ and $10^{-9}$ on timescales of up to $10^6$ years \citep{Garrod2006,Geppert2006}.

In the context of YSOs and their outflows, we assume
and expect all of the grain-surface formation of H$_2$O and CH$_3$OH
to have already taken place during the preceding dark-cloud and prestellar core phases. Material is liberated from the grain surfaces either through thermal desorption
if the temperature of the dust grains rises above $\sim$100 K for water
\citep[][]{Fraser2001, Brown2007} and $\sim$85\,K for methanol \citep[][]{Brown2007} as they journey close to the protostar \citep[][]{Visser2009},
or through sputtering in shocks.
While photodesorption due to cosmic-ray induced UV radiation is
expected to be important in the cold conditions found in protostellar envelopes
\citep[e.g.][Schmalzl et al. in prep.]{Caselli2012,Mottram2013},
it is unlikely to be dominant in the conditions found in outflows.

Therefore, any gas-phase CH$_3$OH observed in outflows will originate from
sputtered ice mantle material, while gas-phase H$_2$O may originate either in warm shocked gas in outflows where the temperature exceeds $\sim$ 200 K \citep{Bergin1999, Charnley1999} or from the grain mantles. \emph{The first objective of this study is to test whether
there is observational evidence in outflow material for water
formation in excess of what we would expect to see from only
ice mantle desorption, i.e. is the gas phase formation route
activated?}

Our method for studying this is
to examine the relative gas-phase column density variations of
CH$_3$OH and H$_2$O derived in components of their emission spectra
associated with the outflow and its shocked cavities, as seen
towards the source positions of three prototypical YSOs. CH$_3$OH
is a particularly good molecule for comparing against H$_2$O
as it is a pure grain mantle species.
Thus, any
variations of $N({\rm CH_3OH})/N({\rm H_2O})$
are a result of independent variations
in either $N({\rm CH_3OH})$ or $N({\rm H_2O})$.

Before we can use $N({\rm CH_3OH})/N({\rm H_2O})$
in studying the variations of $N({\rm H_2O})$, we must first consider whether any CH$_3$OH is destroyed, either by dissociative desorption during sputtering or in the shock through reactions with H. This latter scenario would be somewhat
similar to that involving another grain-surface product,
NH$_3$, at positions offset from the YSO \citep{Codella2010,Viti2011}.
\emph{The second objective of this paper is to provide
observational limits on the extent of CH$_3$OH destruction in the outflows of low-mass YSOs.}
This is accomplished by comparing the column densities to that of a CO line which traces the same gas as H$_2$O, such as CO J=10$-$9 \citep{Yldz2013,Tafalla2013,Santangelo2013,Nisini2013}. The $N({\rm CH_3OH})/N({\rm CO})$ ratio can then be translated
into the CH$_3$OH abundance
$X({\rm CH_3OH}) = N({\rm CH_3OH})/N({\rm H_2})$ by assuming
a constant CO abundance in the outflow, $10^{-4}$.

This paper is structured as follows. In Section
\ref{sec:observations} we describe our observations and
then describe the immediate results in Section \ref{sec:resuls}.
In Section \ref{sec:analysis}
we describe the decomposition of the acquired spectra
into components, and explain which of these we assume to have the same
physical origin. We also discuss the likely physical
conditions in the outflow, and use them to constrain
the radiative transfer
simulations which allow us to convert our observed line
intensity ratios into column density ratios. In Section
\ref{sec:discussion} we discuss the acquired column density
ratios and place them in the context of the arguments and
processes outlined above.
Finally, in Section \ref{sec:conclusions} we summarise
our results and draw our conclusions.

\section{Observations}
\label{sec:observations}

The data used in this paper consist of spectroscopic observations of H$_{2}$O,
CH$_{3}$OH and CO towards three Class 0 protostars: NGC 1333 IRAS 2A, 4A, and 4B. The details
of this sample are given in Table~\ref{tab:observations_sources}. Observations
of the 2$_{02}-$1$_{11}$ para-H$_{2}$O and CO $J$=10$-$9 transitions were obtained with the
Heterodyne Instrument for the Far-Infrared \citep[HIFI;][]{deGraauw2010} on
the \textit{Herschel} Space Observatory \citep{Pilbratt2010} as part of the
``Water in star-forming regions with \textit{Herschel}'' key programme
\citep[WISH;][]{vanDishoeck2011}.
The reduction of these data is discussed further in \citet{Kristensen2010} and \citet{Yldz2013} respectively. The
$J$=5$-$4 and 7$-$6 $K$-ladders of CH$_{3}$OH were observed with the
James Clerk Maxwell Telescope\footnote{The James Clerk Maxwell Telescope is operated by the Joint
Astronomy Centre on behalf of the Science and Technology Facilities Council of
the United Kingdom, the National Research Council of Canada, and (until 31 March
2013) the Netherlands Organisation for Scientific Research.} (JCMT) with the RxA and
RxB receivers, respectively. Details can be
found in \citet{Maret2005}. A summary of the frequencies, beam sizes and upper level energies
($E_{\mathrm{up}}$) can be found in Table~\ref{tab:observations_lines}. A linear baseline was subtracted from all spectra,
which were then placed on a common velocity
scale, centred at the local rest velocity of the source. As the A-CH$_3$OH $7_0-6_0$ data were obtained in a smaller beam than the other transitions, the line intensity was scaled down by a factor of $\sim 1.3-1.5$ to take the difference in beam size into account (see Appendix \ref{app:beamdilution} for further details).

The uncertainties reported in the rest of this paper only relate to the intrinsic random noise in each spectrum. Typical $\sigma_{RMS}$ levels (in channel sizes binned to 1 \kms ; see below) for the H$_{2}$O, CO and two CH$_{3}$OH lines are 0.02, 0.17, 0.15 and 0.24\,K respectively. The calibration uncertainty of the HIFI spectra is $\la$ 10\% \citep{Roelfsema2012} and the calibration uncertainty of the JCMT spectra is $\sim$ 20\%.

\begin{table}
\begin{center}
\caption[]{Observed line parameters.}
\begin{tabular}{lllll}
\hline
\hline \noalign {\smallskip}
Line	& Rest Freq.$\rm ^{(a)}$	& $\theta_{mb}$ & $E_{\mathrm{up}}/k_{\mathrm{b}}$ & $n_{\rm cr}^{\rm (b)}$ \\
	& (GHz) & (\arcsec) & (K) & (cm$^{-3}$) \\
\hline\noalign {\smallskip}
H$_{2}$O $2_{02}-1_{11}$		& $987.927$	& 21$\rm ^{(c)}$			& $100.8$	& 8$\times$10$^7$ \\
CH$_{3}$OH $5_{K}-4_{K}$	& $241-242$	& 21\phantom{$\rm ^{(c)}$}	& $41-136$	&  5$\times$10$^5$ \\
CH$_{3}$OH $7_{K}-6_{K}$	& $338-339$	& 14\phantom{$\rm ^{(c)}$}	& $65-256$	&  1$\times$10$^6$ \\
CO $10-9$					& $1151.985$	& 18$\rm ^{(c)}$			& $304.2$	&  5$\times$10$^5$ \\
\hline\noalign {\smallskip}
\multicolumn{5}{p{.95\linewidth}}{\footnotesize $\rm ^{(a)}$\cite{JPLcat}}\\
\multicolumn{5}{p{.95\linewidth}}{\footnotesize $\rm ^{(b)}$ Critical density at 200 K; for the CH$_3$OH transitions the values reported are for the highest S/N $K$=0 A transitions.} \\
\multicolumn{5}{p{.95\linewidth}}{\footnotesize $\rm ^{(c)}$Calculated using equation 3 from \citet{Roelfsema2012}.}
\label{tab:observations_lines}
\end{tabular}
\end{center}
\end{table}

\begin{table*}
 \centering
 \begin{minipage}{140mm}
\caption[]{Source properties}
\begin{tabular}{llllcccc}
\hline
\hline \noalign {\smallskip}
Source			& \multicolumn{1}{c}{$D$}		& \multicolumn{2}{c}{Coordinates}
				& $\varv_{\rm LSR}^{\rm (a)}$	& $L_{\rm bol}^{\rm (b)}$
				& $T_{\rm bol}^{\rm (b)}$		& $M_{\rm env}^{\rm (c)}$
				\\
				& (pc) 							& $\rm \alpha_{J2000}$
												& $\rm \delta_{J2000}$
				& (\kms)						& ($L_{\sun}$)
				& (K)							& ($M_{\sun}$)
				\\
\hline\noalign {\smallskip}
NGC1333 IRAS2A	& 235 							& $03^{\rm h}28^{\rm m}55\fs 60$
												& $+31^\circ 14'37\farcs 1$
				& $+$7.7						& 35.7
				& \phantom{1}50					& 5.1
				\\
NGC1333 IRAS4A 	& 235							& $03^{\rm h}29^{\rm m}10\fs 50$
												& $+31^\circ 13'30\farcs 9$
				& $+$7.2						& \phantom{1}9.1
				& \phantom{1}33					& 5.2
				\\
NGC1333 IRAS4B	& 235							& $03^{\rm h}29^{\rm m}12\fs 00$
												& $+31^\circ 13' 08\farcs 1$
				& $+$7.1						& \phantom{1}4.4
				& \phantom{1}28					& 3.0
				\\
\hline\noalign {\smallskip}
\multicolumn{8}{p{.95\linewidth}}{\footnotesize $^{\rm (a)}$ Obtained from ground-based C$^{18}$O or C$^{17}$O observations \citep{Yldz2013} with the exception of IRAS4A for which the value from \citet{Kristensen2012} is more consistent with our data.} \\
\multicolumn{8}{p{.95\linewidth}}{\footnotesize $^{\rm (b)}$ Measured using \textit{Herschel}-PACS data from the WISH key programme \citep[][]{Karska2013}.} \\
\multicolumn{8}{p{.95\linewidth}}{\footnotesize $^{\rm (c)}$ Mass within the 10\,K radius, determined by \citet{Kristensen2012} from DUSTY modelling of the sources.} \\
\label{tab:observations_sources}
\end{tabular}
\end{minipage}
\end{table*}

In the upcoming sections, when discussing the intensity ratios
of the various lines included in this study, they have been
resampled to fit a common velocity resolution of 1 \kms\
which was found to give a good balance between resolution
and signal-to-noise ratio (S/N). Resampling also improves the S/N of the
line wings, which are crucial for this study.

\section{Results}
\label{sec:resuls}

\begin{figure*}
 \includegraphics[width=0.95\linewidth]{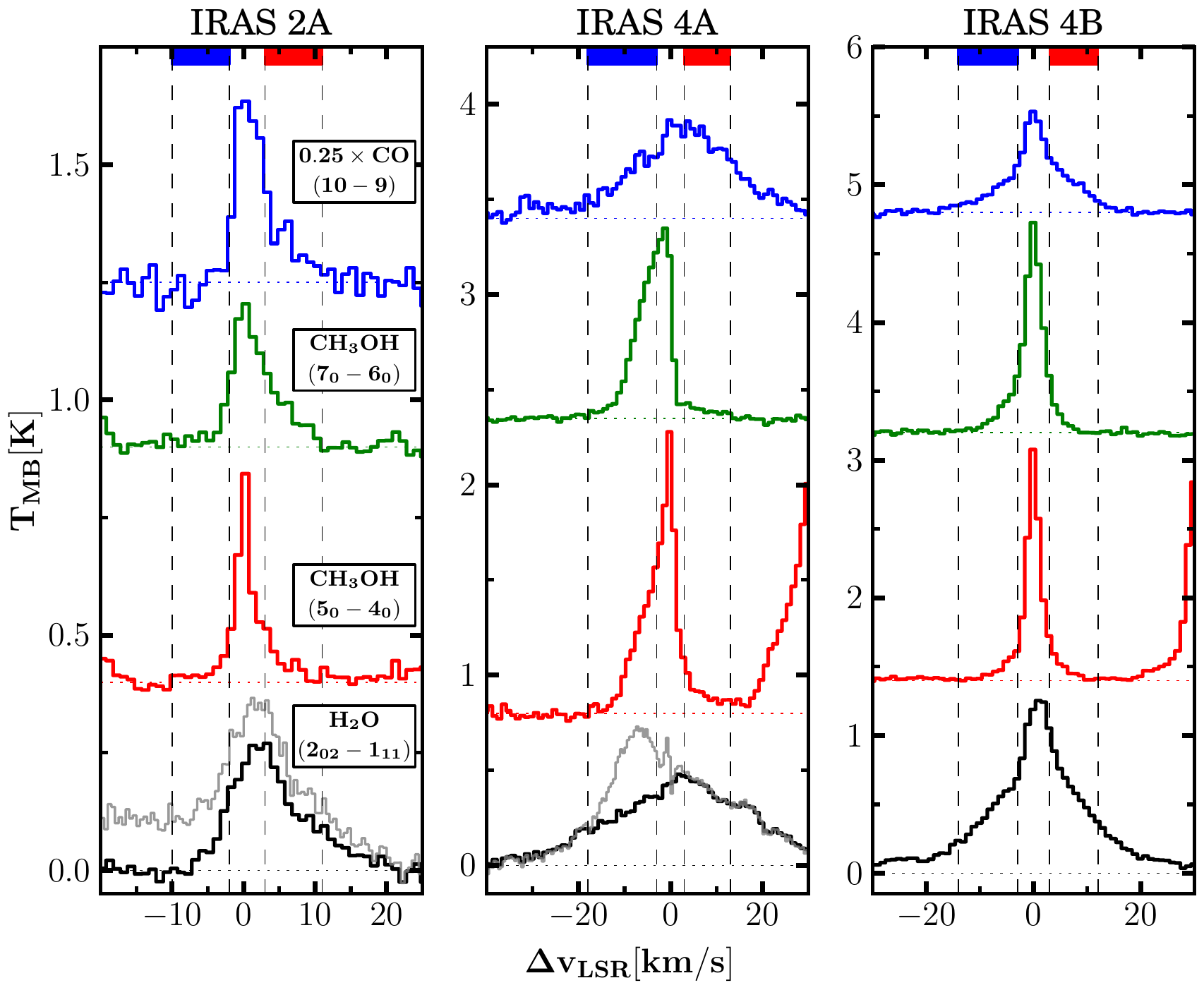}
 \caption{The emission of the $5_{0}-4_{0}$ (red) and $7_{0}-6_{0}$ (green) A-CH$_3$OH, the 10-9 (blue) CO spectra plotted above the $2_{02}-1_{11}$ p-H$_2$O spectrum (gray), which has been decomposed in the manner discussed in Section \ref{sec:velstruct}; decomposed spectra are shown in black. All spectra have been resampled to a common channel size of 1 \kms , and shifted to 0 \kms. The CO 10-9 spectrum has been scaled by 0.25. The spectra are offset along the y-axis for clarity. The dashed vertical lines show the velocity ranges used for selecting the red and blue wings of the spectra, as colour-coded by the horizontal lines at the top.}
 \label{fig:decomp_h2o}
\end{figure*}


\subsection{Line profiles}

Figure \ref{fig:decomp_h2o} shows the H$_2$O spectra of IRAS 2A, IRAS 4A, and IRAS 4B at the source position, overlaid with both of the CH$_3$OH and CO spectra. For the wings of IRAS 2A all profiles are very similar, although the red wing of the H$_2$O spectrum extends to higher velocities than any of the other species by about 5 \kms. IRAS 4A is characterised by the highly asymmetric profile of the CH$_3$OH 7-6 spectrum compared to the H$_2$O and CO 10$-$9 spectra. The H$_2$O line wings towards both IRAS 4A and 4B extend to significantly higher velocities than the CH$_3$OH and CO line wings by more than 10 \kms.

Previous studies have found that the observed H$_2$O line profiles are generally complex and consist of multiple dynamic components \nocite{Kristensen2010b, Kristensen2011, Kristensen2012, Kristensen2013}\citep[Kristensen et al. 2010b, 2011, 2012, 2013;][]{Mottram2013}. These components are typically associated with the envelope, outflowing and shocked gas and trace physical components not detected previously. CH$_3$OH profiles, on the other hand, typically consist of just two components: one associated with the outflow and one with the envelope. In the following we limit ourselves to studying the line wings only, and ignore the line centers which are typically associated with envelope emission.

The velocity ranges used as the red and blue wings of the objects are indicated by the respectively coloured bars at the top of Fig. \ref{fig:decomp_h2o}. The ranges were selected such that any envelope contribution is excluded (typically $\pm$ 2 \kms\ from the source velocity) and such that the line wings are completely included. Towards IRAS4A and 4B, the red-shifted A-CH$_3$OH $5_0-4_0$ wing emission starts to blend with E-CH$_3$OH $5_{-1}-4_{-1}$ emission at velocities of $+$ 5$-$10 \kms\ from the source velocity, which therefore sets the upper limit on the red-shifted velocity.

\subsection{Observed line ratios}
\label{sec:ratios}
\begin{figure*}
 \includegraphics[width=0.95\linewidth]{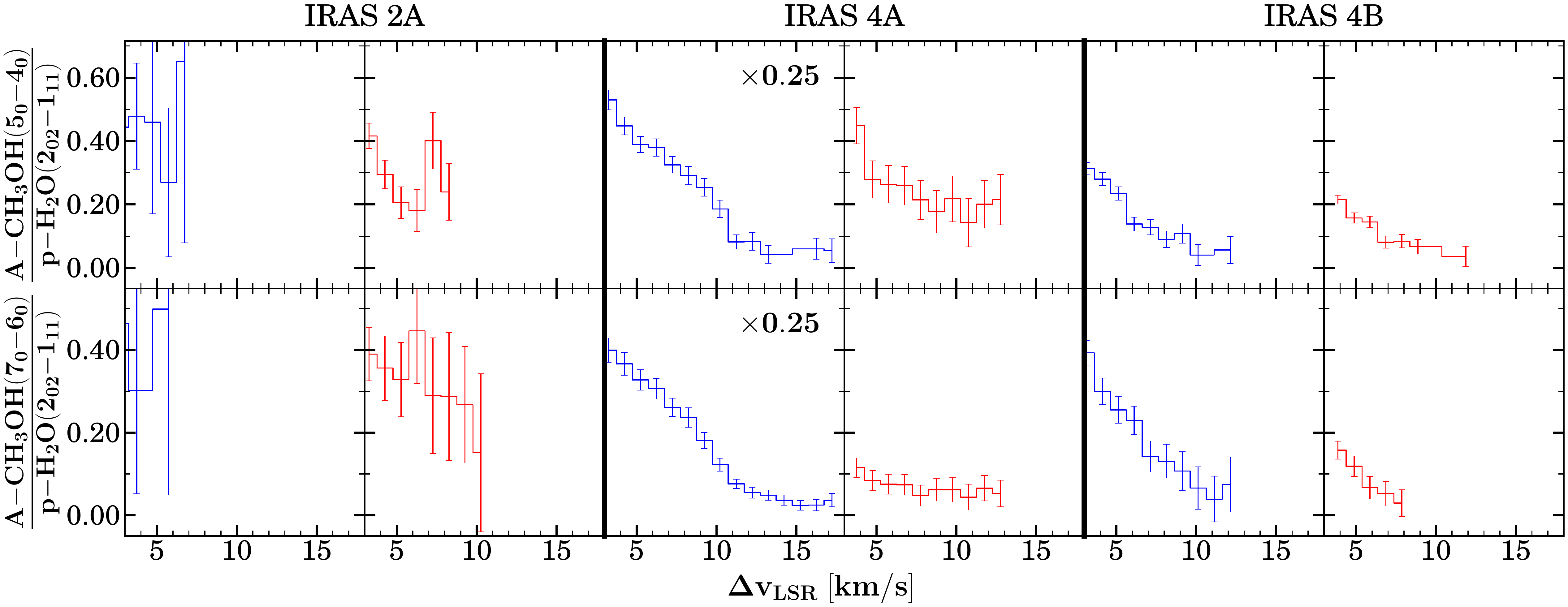}
 \includegraphics[width=0.95\linewidth]{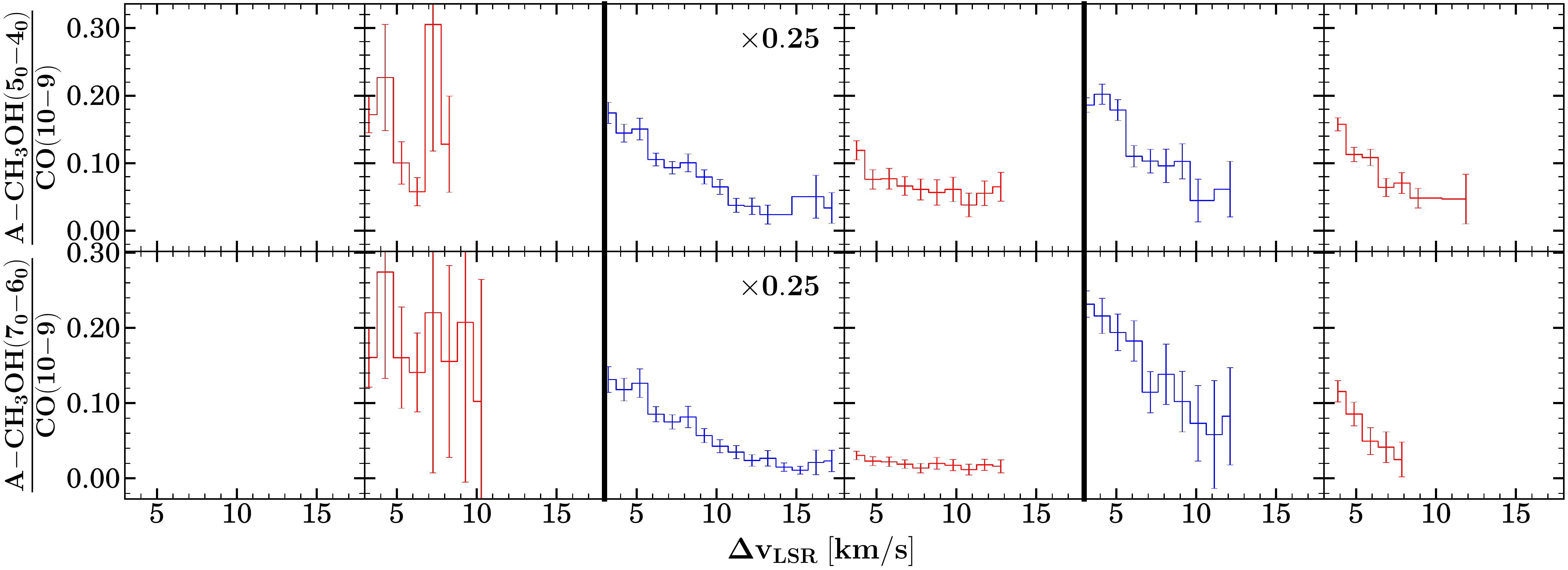}
 \caption{The ratio of the line intensities of two different A-CH$_3$OH lines ($5_0-4_0$ and $7_0-6_0$ over the line intensity of the $2_{02}-1_{11}$ p-H$_2$O transition (\textit{top}), CO 10$-$9 (\textit{bottom}). The velocity along the x axis is absolute offset with respect to the source velocity for the blue and red wings, respectively. Only the velocity range covering the line wings tracing emission related to the outflows is included in the line ratios and all points where S/N $<$ 1 for either line have been ignored. Some of the line ratios were scaled for visibility reasons (marked). The error bars only include measurement uncertainties and not relative calibration uncertainties. }
 \label{fig:div_h2o}
\end{figure*}

The line ratios of the two ground-state 5$-$4 and 7$-$6 A-CH$_3$OH transitions with respect to H$_2$O $2_{02}-1_{11}$ and CO 10$-$9 are presented in Fig. \ref{fig:div_h2o}. The line ratios are measured where $\Delta v \geq 3$ \kms (at lower velocities we can not separate outflow emission from quiescent envelope emission), and where the S/N ratio is $>$ 1 in both transitions. The CO 10$-$9 line profile towards IRAS2A shows no blue-shifted emission in spite of the high quality of the data.

The CH$_3$OH / H$_2$O line ratios show a decreasing trend with increasing velocity over the range where CH$_3$OH is detected. This is particularly visible towards IRAS4A in the blue wing, where the ratio decreases by more than a factor of 5. Most other line wings show a decrease of a factor of 2$-$3. Only the IRAS2A line wings appear to exhibit constant line ratios, although the higher-S/N red $7_0-6_0$ CH$_3$OH line wing shows a significant decrease with respect to H$_2$O emission. The CH$_3$OH / CO 10$-$9 line ratios (Fig. \ref{fig:div_h2o}) exhibit similar trends as the ratios with H$_2$O, consistent with the similarity of the H$_2$O and CO 10$-$9 profiles \citep{Yldz2013}.

\section{Analysis}
\label{sec:analysis}

\subsection{\textsc{RADEX} and the degeneracy of the parameter space}
\label{sec:radex}


The non-LTE excitation and radiative transfer software \textsc{RADEX} \citep{VanderTak2007} is used to simulate the physical
conditions of a body of gas in order to convert integrated molecular line intensities into
corresponding column densities. The free parameters are the gas density ($n({\rm H_2})$), temperature
($T$) and either the molecular column density ($N$) or the opacity ($\tau$), given a linewidth $\Delta v$. \textsc{RADEX} relies on the escape probability approximation and assumes a slab geometry. In all following simulations except
those used for generating Figure \ref{fig:radex_water4} (where line width is used to control $\tau$)
we set the line width to 1 \kms, corresponding to the channel width. In the following we assume that
the H$_2$O o/p ratio is 3.0 \citep{Emprechtinger2013}. All column densities involving H$_2$O presented for the rest of the paper have made use of this ratio to convert $N$(p-H$_2$O) to $N$(H$_2$O). The molecular line information \textsc{RADEX} uses is
taken from LAMDA, \citep{Schoier2005} with information for the individual molecules from
\citet{Rabli2010}, \citet{Yang2010} and \citet{Daniel2011}.

Constraints must be placed on the parameter space if we are to obtain a non-degenerate solution to our data analysis. Since we are only interested in the relative abundance variations of CH$_3$OH, CO and H$_2$O, the line intensity ratios must be converted to column density ratios. In the following subsections we discuss what constraints can be put on the gas density, temperature and opacity based on relevant envelope and outflow parameters, and what the effect of varying these parameters is on the final column density ratio.

\subsubsection{Opacity and column density}

From \citet{Yldz2013} we know that the CO 10$-$9 line wings are optically thin. For CH$_3$OH, observations towards pure outflow positions, such as L1157 B1 and B2 \citep{Bachiller1995}, result in inferred column densities in the line wings that suggest they are optically thin. We assume that this is true towards the protostellar position as well.

Water emission is known to be optically thick towards the protostellar position \citep{Kristensen2011} with opacities $>$ 10 for the ground-state 557 GHz $1_{10}-1_{01}$ transition. However, because of the high critical density of water ($>$ 10$^8$ cm$^{-3}$ for the 987 GHz transition studied here) compared to the H$_2$ density, the effective critical density is $A$/($C\tau$) where $A$ is the Einstein A coefficient, $C$ is the collisional rate coefficient and $\tau$ is the opacity. In the limit where the opacity is high but $n < n_{\rm crit}$, the emission becomes effectively thin and a photon is lost for each radiative decay \citep{Snell2000}, i.e. while the emission may be optically thick it is effectively optically thin.

\begin{figure}
 \includegraphics[width=0.95\linewidth]{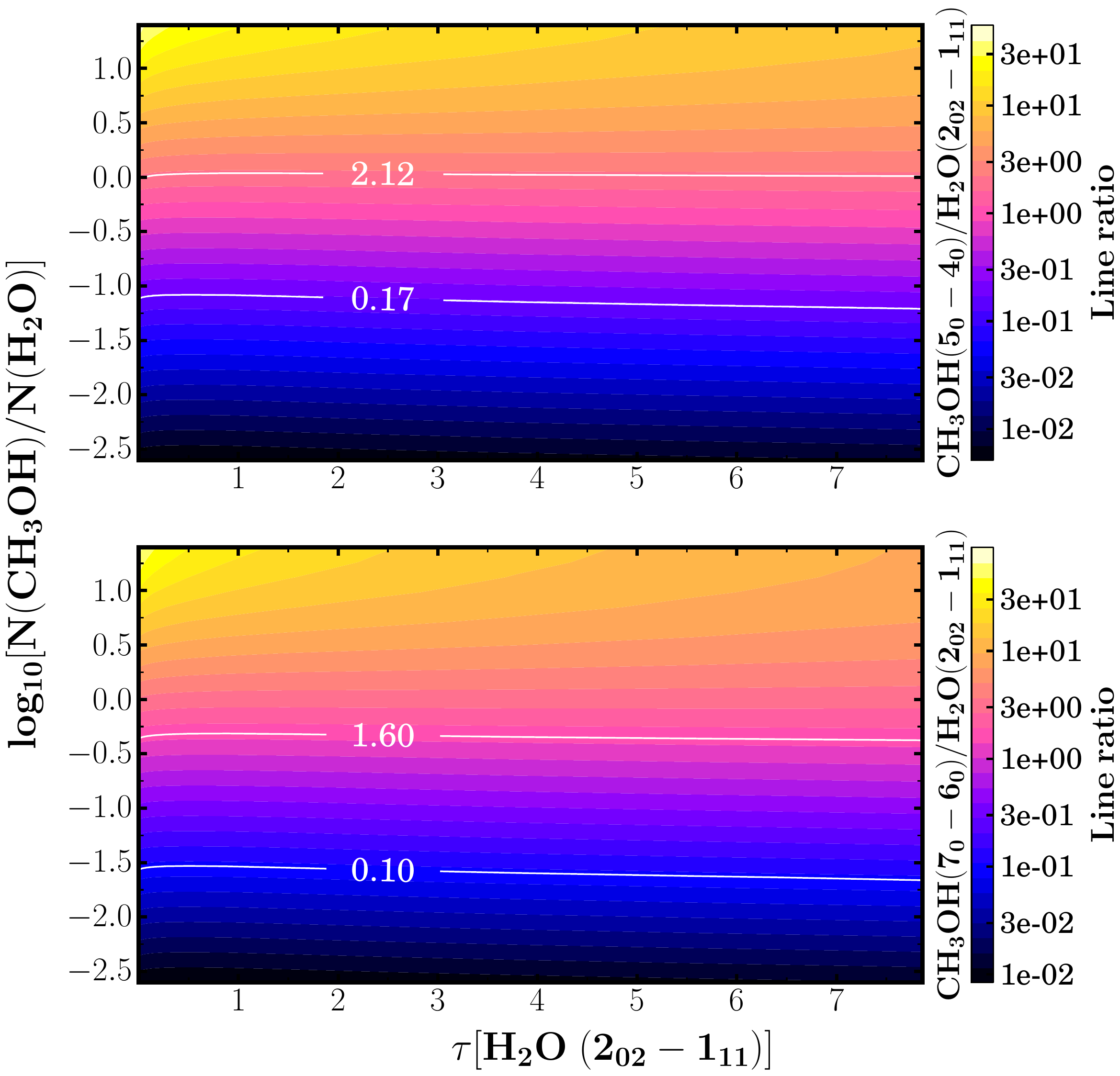}
 \caption{A visualization of \textsc{RADEX} simulations which show
 how the conversion of line intensity ratio to column density
 ratio behaves as a function of the optical depth of the H$_2$O
 emission in the case of CH$_3$OH versus H$_2$O, for two different
 CH$_3$OH transitions. The white
 contour lines show the maximum and minimum line intensity ratios
 the blue wing of IRAS 4A. $T_{\rm kin}$ is fixed at 200 K, and
 $n({\rm H_2})$ is fixed to be $10^6\ {\rm cm^{-3}}$.}
 \label{fig:radex_water4}
\end{figure}

A grid of \textsc{RADEX} simulations was run to study how increasing opacity affects the conversion of the observed line intensity ratios into column density ratios (Fig. \ref{fig:radex_water4}). As can be seen, the effects of varying the opacity are limited, much less than a factor of two. The reason is that although the water emission is optically thick, it is effectively thin (see above). In the following we therefore assume that the water emission is optically thin.

An additional constraint can be put on the CH$_3$OH/H$_2$O column density ratio. If the relative gas phase abundance of CH$_3$OH and H$_2$O trace the relative ice phase abundance of the same molecules, values are expected to range from a few to 30\% \citep[e.g.][]{Dartois1999, Pontoppidan2003, Pontoppidan2004, Gibb2004, Boogert2008, Oberg2011}. Should gas-phase water formation be taking place, the gas-phase abundance ratio would be lower than the ice phase abundance ratio, and $[N({\rm CH_3OH})/N({\rm H_2O})]_{\rm gas} < [N({\rm CH_3OH})/N({\rm H_2O})]_{\rm ice}$. Using $T_{\rm kin}=200\ {\rm K}$ and $n({\rm H_2}) = 10^{6}\ {\rm cm^{-3}}$ ensures that the inferred $N({\rm CH_3OH}) < N({\rm H_2O})$ is always true for our data, even with optical depth effects accounted for.

\begin{table}
\caption{Physical conditions in the \textsc{RADEX} simulations}
\begin{tabular}{llll}
\hline
\hline \noalign {\smallskip}
Parameter			& Value 						& Range
					& Effect						\\
\hline
$\tau ({\rm H_2O})$ & Optically thin$^{\rm (a)}$	& $0 - 20$
					& $<$0.1 dex					\\
$T_{\rm kin}$		& $200\ {\rm K}$				& $100 - 300\ {\rm K}$
					& 0.5 dex						\\
$n({\rm H_2})$		& $10^{6}\ {\rm cm^{-3}}$		& $10^{5} - 10^{7}\ {\rm cm^{-3}}$ 
					& 1.0-1.5 dex$^{\rm (b)}$		\\
\hline
\multicolumn{4}{p{.95\linewidth}}{\footnotesize $\rm ^{(a)}$ The H$_2$O emission is optically thick but effectively thin \citep[see text and][]{Snell2000}.}\\
\multicolumn{4}{p{.95\linewidth}}{\footnotesize $\rm ^{(b)}$ $n({\rm H_2})$ has a greater effect on $N$(CH$_3$OH)/$N$(H$_2$O) than on $N$(CH$_3$OH)/$N$(CO)}\\
\label{tab:physcond}
\end{tabular}
\end{table}

\subsubsection{Temperature}

Examining the rotational temperatures (presented separately in Appendix \ref{app:rotdia}) of CH$_3$OH towards our objects places them above "normal" cloud temperatures of 10-15 K but below outflow temperatures of 100$-$300 K. Though CH$_3$OH is not necessarily in LTE, it is not far off and we proceed with assuming that LTE applies for the CH$_3$OH lines, i.e. that $T_{\rm kin}=T_{\rm rot}$.
The kinetic temperature range used in the \textsc{RADEX} simulations is 100$-$300 K, based on the universal warm temperature component seen in CO rotational diagrams of low-mass protostars \citep{Herzceg2012,Karska2013,Manoj2013,Green2013}.

\begin{figure}
 \includegraphics[width=0.95\linewidth]{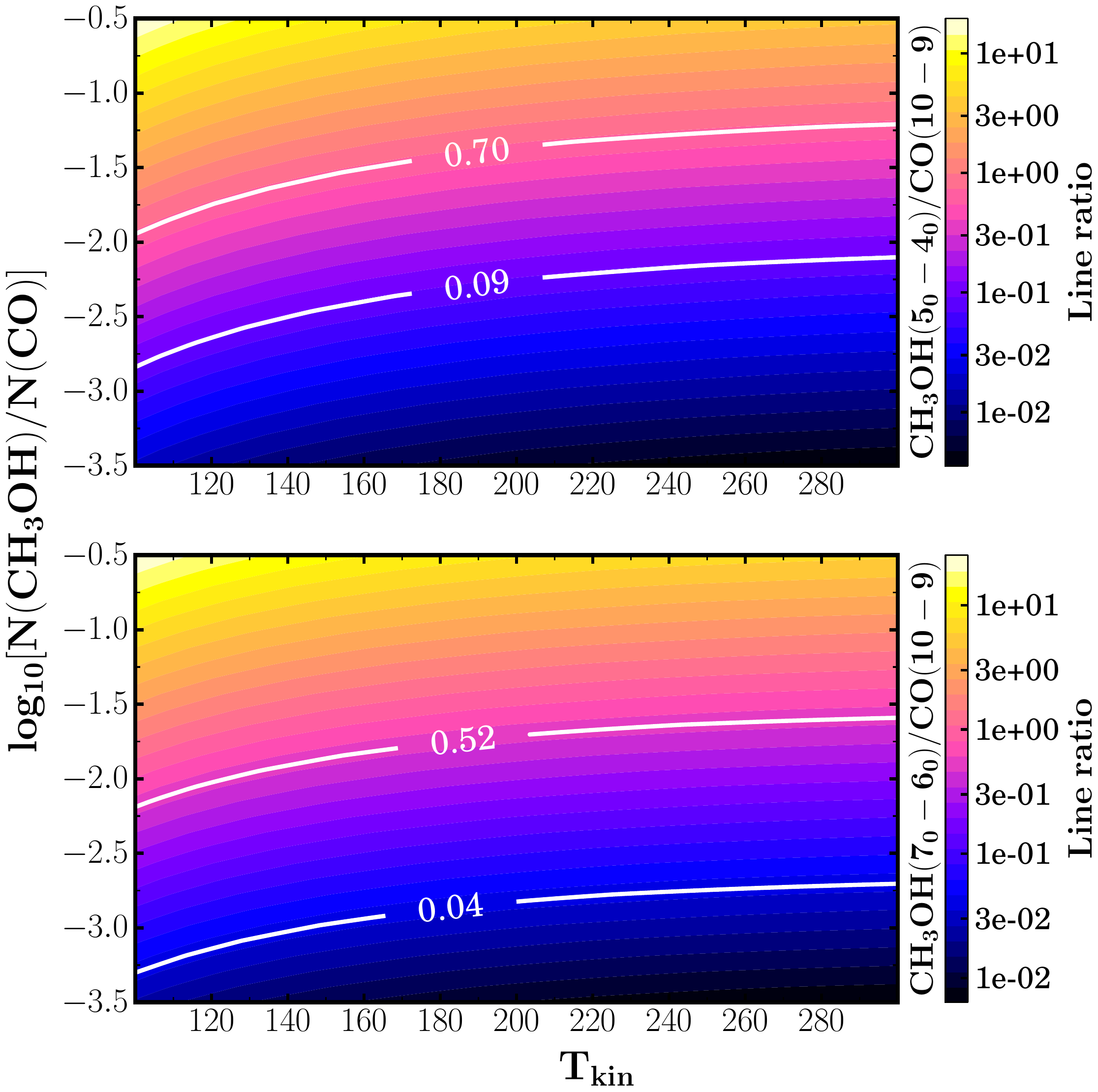}
 \caption{The effects of varying $T_{\rm kin}$ on the derived
 $N({\rm CH_3OH})/N({\rm CO})$ ratio in the optically thin limit for
 $n({\rm H_2})=10^{6}\ {\rm cm^{-3}}$, for two different
 CH$_3$OH transitions. The
 white contour lines show the maximum and minimum
 observed line intensity ratios for the blue wing of
 IRAS 4A, as shown in Fig. \ref{fig:div_h2o}.}
 \label{fig:radex_methanol2}
\end{figure}

\begin{figure}
 \includegraphics[width=0.95\linewidth]{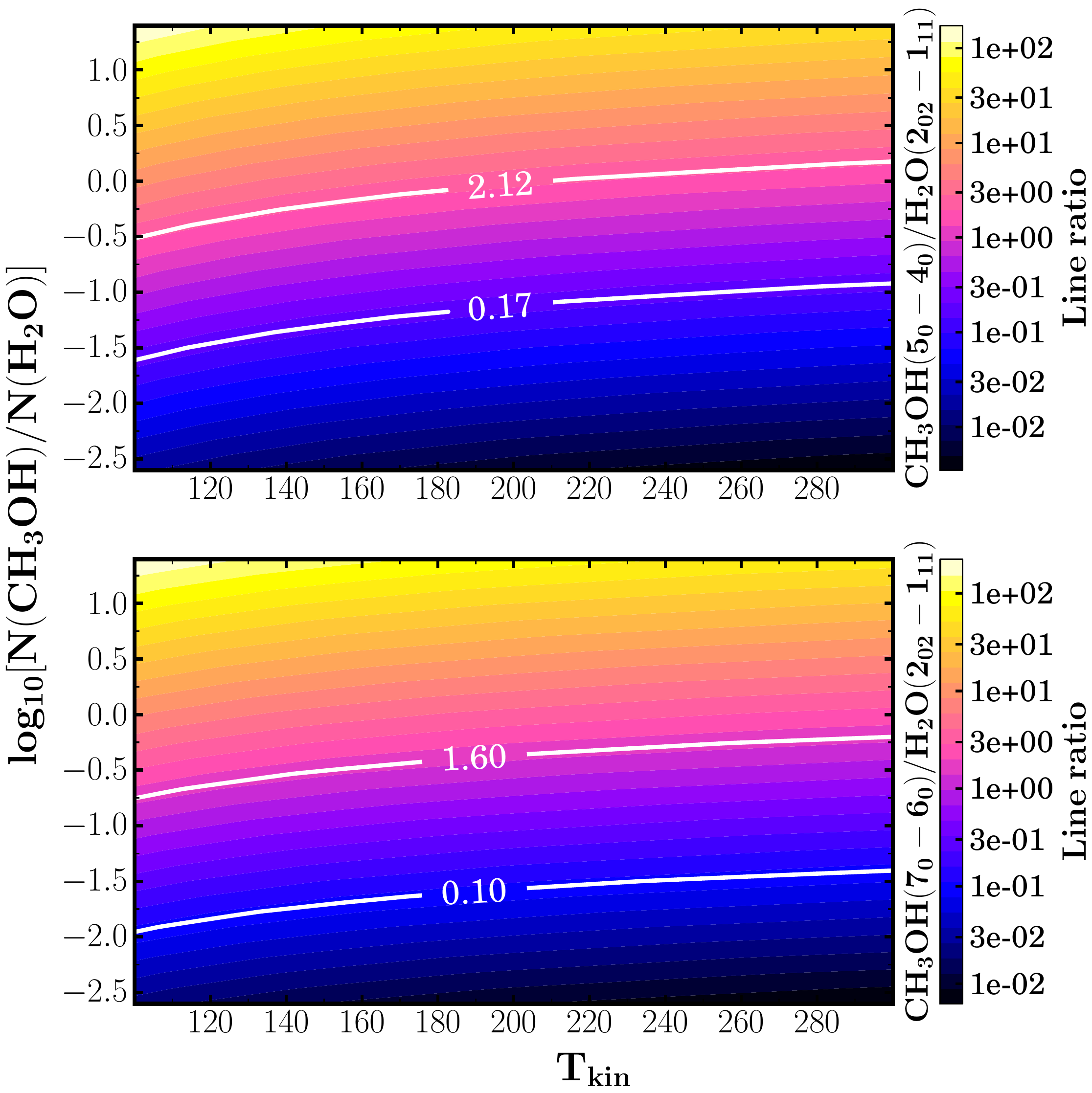}
 \caption{The effects of varying $T_{\rm kin}$ on the derived
 $N({\rm CH_3OH})/N({\rm H_2O})$ ratio in the optically thin limit,
for $n({\rm H_2})=10^{6}$ cm$^{-3}$, for two different
 CH$_3$OH transitions. The
 white contour lines show the maximum and minimum
 observed line intensity ratios for the blue wing of
 IRAS 4A, as shown in Fig. \ref{fig:div_h2o}.}
 \label{fig:radex_water2}
\end{figure}

Figure \ref{fig:radex_methanol2} shows that the inferred CH$_3$OH/CO column density ratio increases by a factor of $\sim$ 3 from 100 to 300 K for a constant density of 10$^6$ cm$^{-3}$. The CH$_3$OH/H$_2$O column density ratio (Fig. \ref{fig:radex_water2}) shows similar behaviour.

\subsubsection{Density}

One way of obtaining estimates of $n({\rm H_2})$ is to use the envelope density at a given distance from the protostar for the outflow material \citep[e.g.][]{Kristensen2012,Yldz2013}. \citet{Kristensen2012} estimated the envelope density as a function of distance for all sources in this sample based on fits to the continuum SED and submm continuum maps, assuming a powerlaw density structure. Densities of $\sim$ 5 $\times$ 10$^6$ cm$^{-3}$ are found at a distance of 1000 AU, corresponding to $\sim$ 4\arcsec\ or half the beam radius. At the edge of the beam (10\arcsec, 2350 AU) the density is typically 10$^6$ cm$^{-3}$. Since the entrained gas in the outflow cavity walls can be either compressed by the shocks or expanding into the outflow cavity, we study $n({\rm H_2})$ in the range of $10^{5} - 10^{7}\ {\rm cm^{-3}}$.

For comparison, the critical densities are listed in Table 2. The envelope densities exceed the CH$_3$OH and CO critical densities, whereas the H$_2$O critical density is never reached. Thus, the CH$_3$OH and CO level populations are likely thermally excited, whereas H$_2$O is sub-thermally excited. For CO and H$_{2}$O this has been shown to be the case by \citet{Herzceg2012} using excitation analysis of \textit{Herschel} PACS observations.

\begin{figure}
 \includegraphics[width=0.95\linewidth]{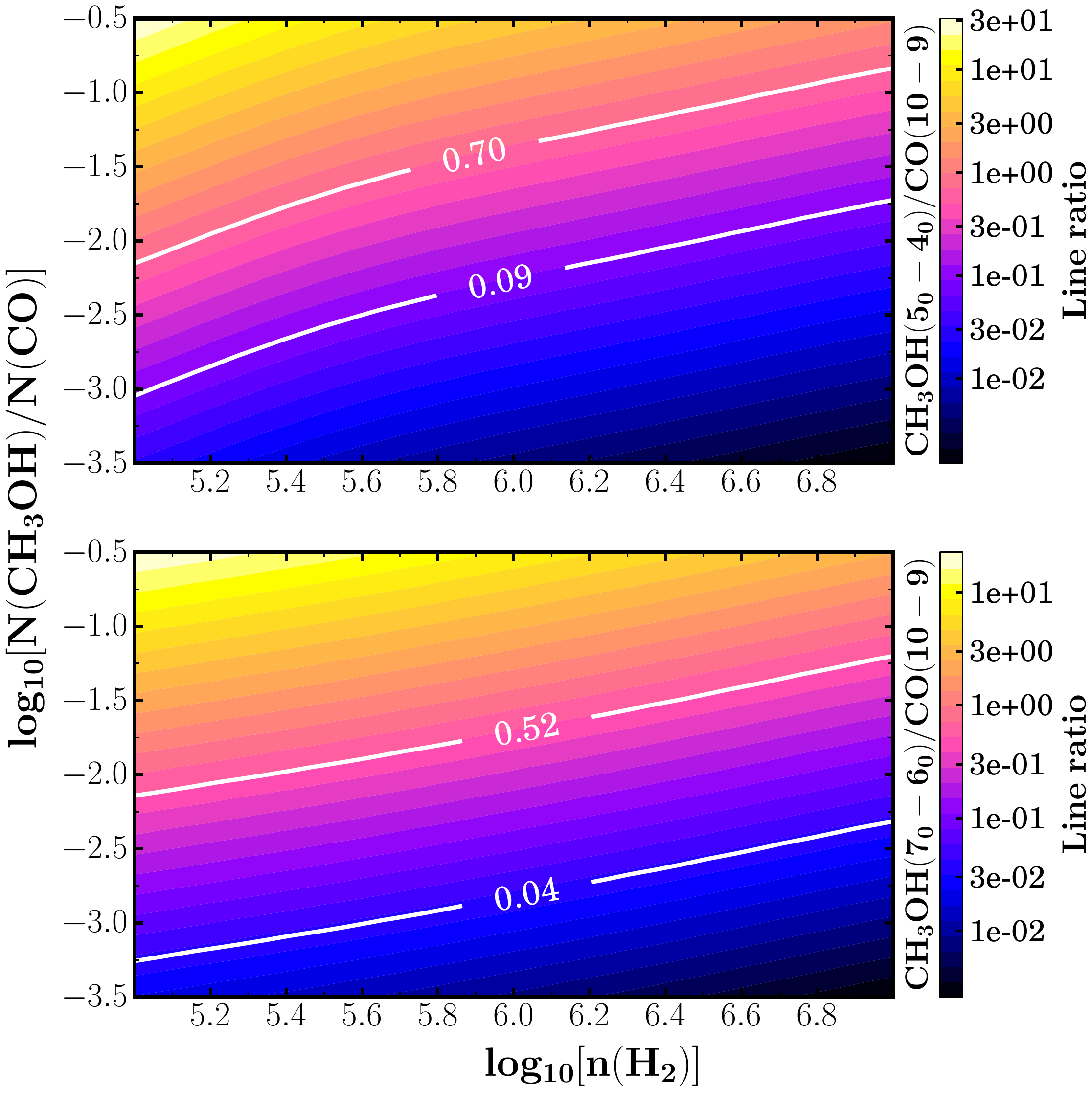}
 \caption{The simulated intensity ratio of CH$_{3}$OH $5_0-4_0$ (top) and
 $7_0-6_0$ (bottom) to CO $10-9$ as contours of $\log_{10}$($n({\rm H_2})$) (x axis) and $\log_{10}$($N({\rm CH_3OH})/N({\rm CO})$) (y axis).
 The
 white contour lines show the maximum and minimum
 observed line intensity ratios for the blue wing of
 IRAS 4A, as shown in Fig. \ref{fig:div_h2o}.
 The contour levels are logarithmically scaled.
 The kinetic temperature is fixed at 200 K and all lines
 are optically thin.}
 \label{fig:radex_methanol1}
\end{figure}

The effect the density range has on the line intensity ratio to column density ratio translation can be examined by running a grid of \textsc{RADEX} simulations in which $\tau$ and $T_{\rm kin}$ are kept constant (optically thin and 200 K, respectively) and $n({\rm H_2})$ and the column density ratio are varied in order to produce line intensity ratios comparable to the observations. As both the CH$_3$OH and CO 10$-$9 line wings are optically thin \citep[][for CO]{Yldz2013}, they form a good benchmark for exploring the effects of the physical environment on the simulated line ratios while allowing us to ignore $\tau$. The results of the \textsc{RADEX} simulation grid are shown as a contour plot in Figure \ref{fig:radex_methanol1}, where they are compared with the observed ratios for the blue wing of IRAS 4A.

\begin{figure}
 \includegraphics[width=0.95\linewidth]{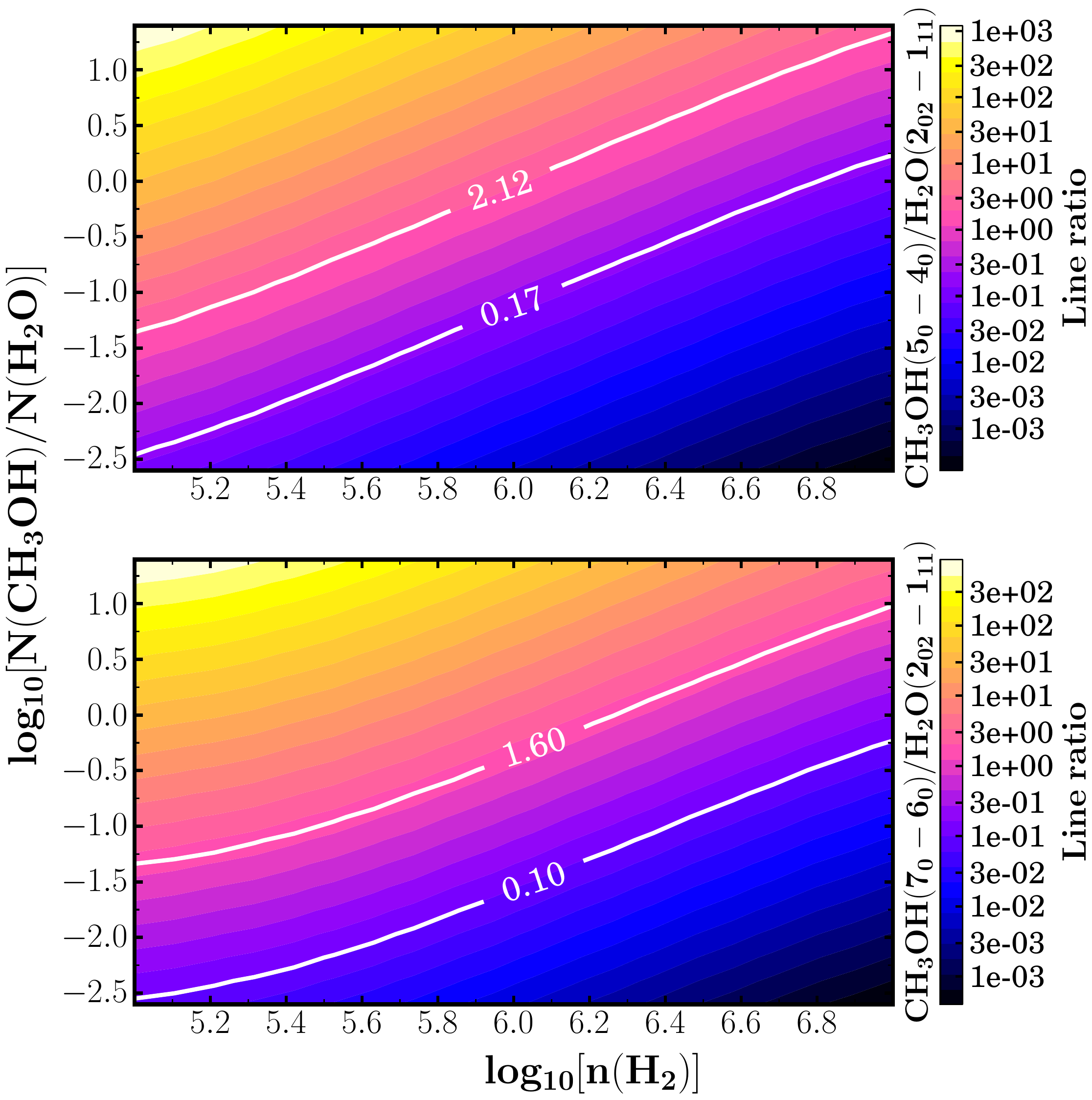}
 \caption{As Figure \ref{fig:radex_methanol1}, but for
 CH$_3$OH versus H$_2$O. $T_{\rm kin}$ is fixed at
 200 K for both species.}
 \label{fig:radex_water3_alt}
\end{figure}

Figure \ref{fig:radex_methanol1} shows that variations of $n({\rm H_2})$ across the assumed range can produce changes in the $N({\rm CH_3OH})/N({\rm CO})$ ratios of a factor of 10, but that the increase depends only weakly on $n$(H$_2$), as is expected when emission is thermalised and optically thin. The same exercise can be done for CH$_3$OH versus H$_2$O and the \textsc{RADEX} results are presented in Figure \ref{fig:radex_water3_alt}. Because H$_2$O excitation is sub-thermal, the effect of $n({\rm H_2})$ on this ratio is stronger than for CH$_3$OH versus CO. The resulting systematic variation in this case is $\pm$1 order of magnitude and varies linearly with density.

\subsection{Molecular column density ratios}
\label{sec:molvar}
With the appropriate range in physical conditions established we are ready to address our original questions by exploring how the observed line intensity ratios convert into column density ratios. We start by examining the CH$_3$OH/CO column density ratio in order to quantify how much CH$_3$OH from the grains reaches the gas phase intact. Second, we examine the CH$_3$OH/H$_2$O ratio. All assumed excitation conditions are listed in Table \ref{tab:physcond} and the resulting column density ratios are summarised in Table \ref{tab:abundance}. The excitation conditions are assumed to be constant as a function of velocity. In order to convert $N$(p-H$_2$O) into $N$(H$_2$O) we have assumed a constant ortho-to-para ratio of 3:1 and thus multiplied all calculated p-H$_2$O column densities by 4.

\begin{figure*}
 \includegraphics[width=0.95\linewidth]{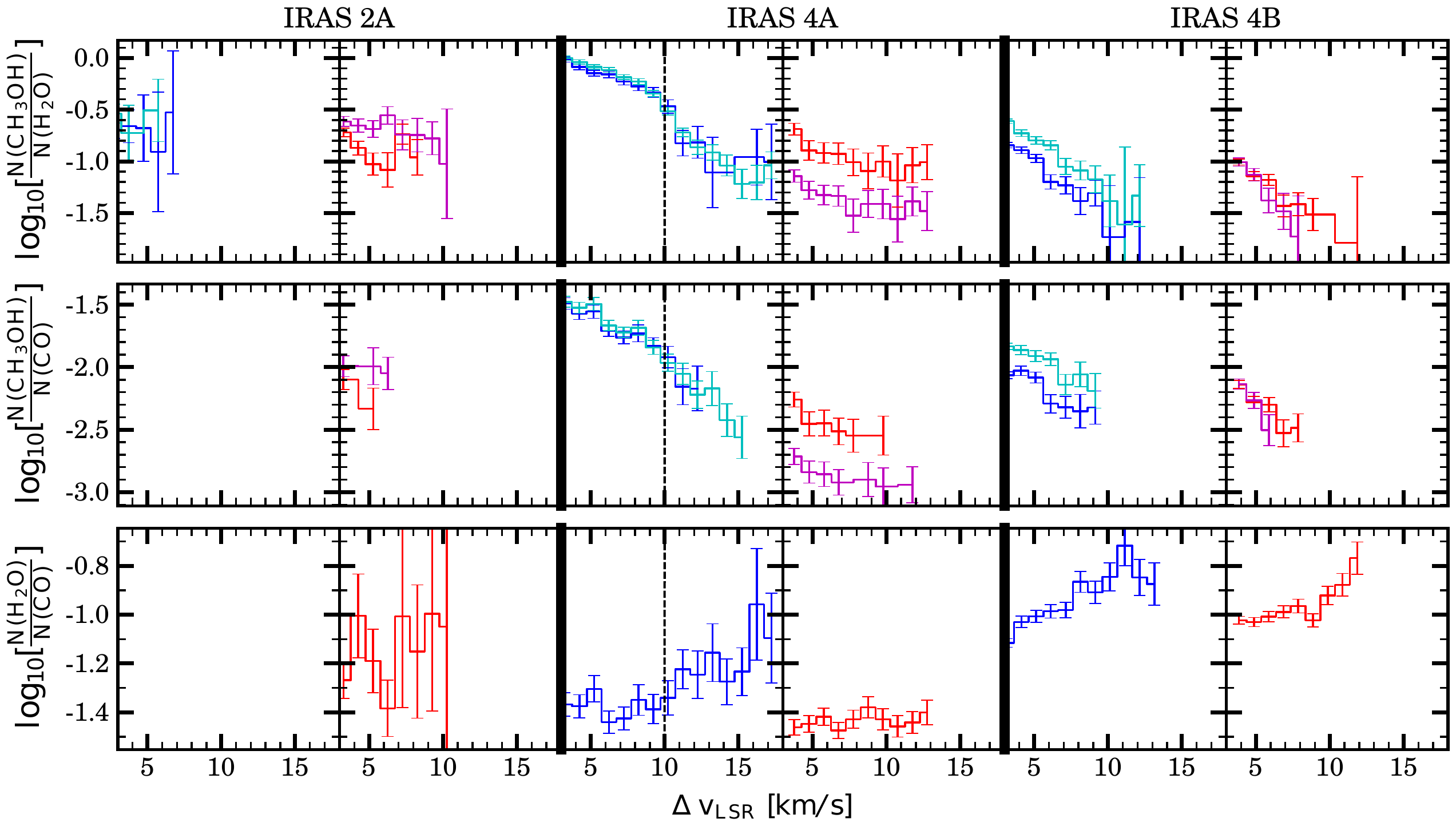}
 \caption{Line intensity ratios converted to corresponding column density ratios using the physical parameters presented in Table \ref{tab:physcond}. The two top rows have had their ratios calculated using both the $5_0-4_0$ (red and blue) and $7_0-6_0$ (cyan and magenta) lines of CH$_3$OH. The dashed line at $\Delta v_{\rm LSR} \approx 10$ \kms in the plots of the blue wing of IRAS 4A represents the approximate velocity where we see a change (compared to lower velocities) in the behaviour of $N({\rm CH_3OH})/N({\rm H_2O})$ and $N({\rm H_2O})/N({\rm CO})$. The line intensity ratios from which the column density ratios of the bottom row are calculated have been presented in \citet{Yldz2013}. Velocity is given with respect to the source velocity.}
 \label{fig:radex_convert}
\end{figure*}

The CH$_3$OH/CO column density ratios are presented in the middle row of Fig. \ref{fig:radex_convert}. Because the line wing emission is assumed optically thin, the column density ratio is proportional to the line ratio. With this assumption, we derive a CH$_3$OH/CO column density ratio that decreases with increasing velocity towards all sources by up to a factor of five (IRAS4A blue). The CH$_3$OH/CO column density ratio is on average $\sim$ 10$^{-3}$.
We follow the example of \cite{Yldz2013} and assume a CO/H$_2$ abundance of 10$^{-4}$. This value may be too high by a factor of a few \citep{Dionatos2013,Santangelo2013}, or too low by a factor of a few \citep{Lacy1994}.
When using this CO abundance the derived CH$_3$OH abundance becomes of the order of 10$^{-7}$, consistent with what is observed towards other outflows \citep[e.g.][]{Bachiller1995, Tafalla2010}.

\begin{table}
\begin{center}
\caption{Column density ratio ranges from the lowest (3 \kms) to highest velocities (10$-$18 \kms, depending on source)$^{\rm (a)}$.}
\begin{tabular}{l c c c c c}
\hline
\hline
Source & \multicolumn{2}{c}{$N$(CH$_3$OH)/$N$(CO)} && \multicolumn{2}{c}{$N$(CH$_3$OH)/$N$(H$_2$O)} \\
& 3 \kms & max$( \varv )$ && 3 \kms & max$( \varv )$ \\
\hline
IRAS 2A - blue 		& \ldots  &\ldots 	&& 2($-$1)	& 2($-$1) \\
IRAS 2A - red 		& 1($-$2) & 1($-$2) && 2($-$1)	& 2($-$1) \\
IRAS 4A - blue 		& 3($-$2) & 3($-$3) && 1(0)		& 8($-$2) \\
IRAS 4A - red 		& 5($-$3) & 3($-$3) && 3($-$1)	& 8($-$2) \\
IRAS 4B - blue 		& 1($-$2) & 4($-$3) && 2($-$1)	& 3($-$2) \\
IRAS 4B - red 		& 8($-$3) & 3($-$3) && 1($-$1)	& 2($-$2) \\
\hline
{\footnotesize $\rm ^{(a)}$ $a$($b$) = $a\times10^b$.}
\label{tab:abundance}
\end{tabular}
\end{center}
\end{table}

The CH$_3$OH/H$_2$O column density ratio decreases with increasing velocity offset, except towards IRAS 2A (Fig. \ref{fig:radex_convert} top row). The average column density ratio for all objects is $\sim$ 10$^{-1}$. The $N({\rm CH_3OH})/N({\rm H_2O})$ ratios calculated using the two different emission lines of CH$_3$OH are noticeably different in the case of the red wing of IRAS 4A, by about a factor of 3, but the two lines show the same trend. Up until 10 \kms the decrease in CH$_3$OH column density with respect to both CO and H$_2$O match each other closely and this is also reflected in the the H$_2$O abundance being close to constant when compared to CO 10$-$9, as is seen in the middle row of Figure \ref{fig:radex_convert}. After 10 \kms this behaviour changes especially towards the blue wing of IRAS 4A and possibly towards the red wing of IRAS 4B as well, with $N$(CH$_3$OH)/$N$(H$_2$O) decreasing slightly more rapidly than $N$(CH$_3$OH)/$N$(CO). The effect is also seen in $N$(H$_2$O)/$N$(CO) which begins increasing at velocities above 10 \kms .

\section{Discussion}
\label{sec:discussion}
\subsection{Velocity structure and the origin of the line emission}
\label{sec:velstruct}
\begin{figure*}
 \includegraphics[width=0.95\linewidth]{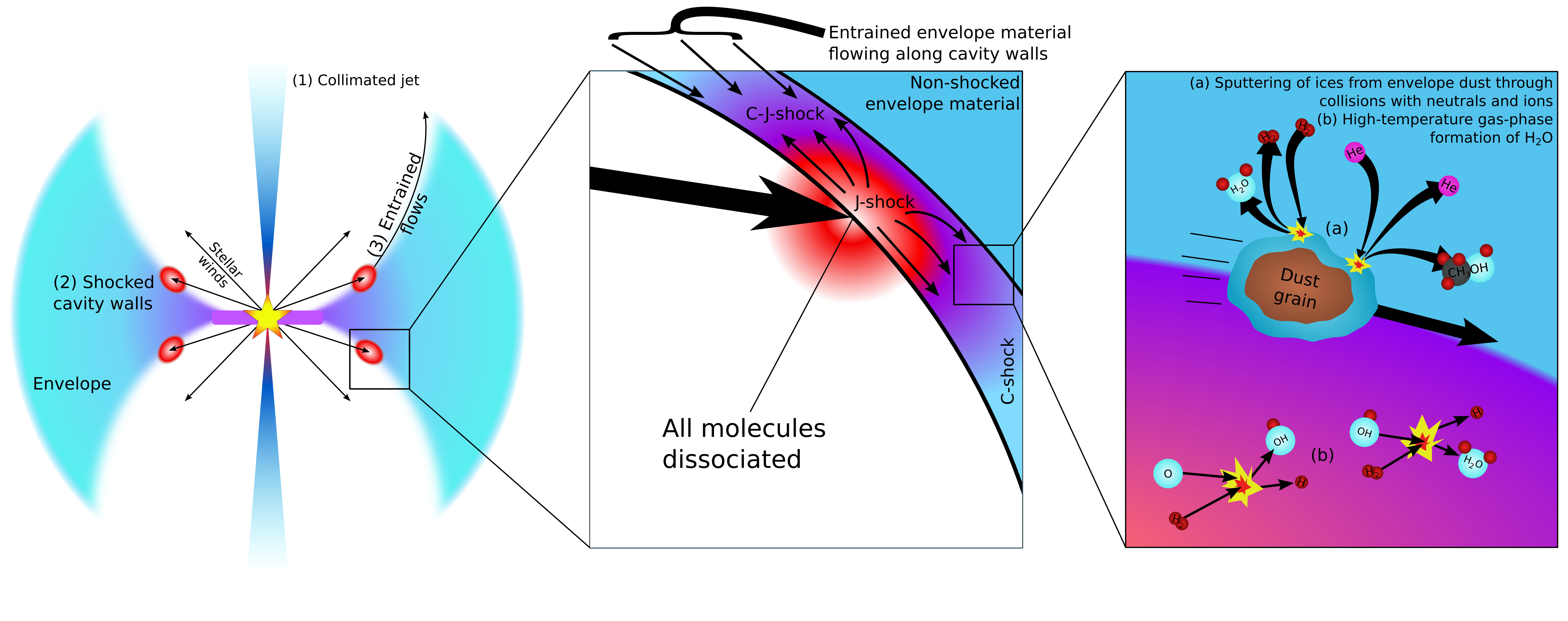}
 \caption{A cartoon (not to scale) illustrating the flow of material in the outflow regions of a young stellar object. The left part shows an overview of the region around the YSO, with the protostar and its protoplanetary disk in the middle of the envelope.
A closer view of one of the shocked regions is illustrated in the middle part. The large arrow represents a maximum point of impact of the stellar wind on the envelope material, where the wind creates a very high-temperature (several $10^4$ K) J-type shock which causes the dissociation of all affected molecules. Further away from the impact point a C-J-type shock (non-steady state shock) is created and even further away a stable C-type shock develops. In addition to the deflected material of the wind flowing along the cavity walls, material from the envelope is entrained and forced to flow along the direction of the deflected wind.
The right figure shows the relevant chemical processes happening
in the shocked material: (a) the sputtering
of ices from grain surfaces by collisions with neutrals and ions, and
(b) the high-temperature gas-phase formation of water.}
 \label{fig:ysowinds}
\end{figure*}

The protostellar wind and jet interact with the envelope in several locations:
\begin{enumerate}
\renewcommand{\theenumi}{(\arabic{enumi})}
\item the stellar wind around the highly collimated bi-polar jet;
\item the shocked material where the stellar winds collide with the outflow cavity walls;
\item the entrained material of the envelope, swept up but not necessarily shocked by the deflected stellar wind material.
\end{enumerate}
These are illustrated in the left part of Figure \ref{fig:ysowinds}.

The last two of these are the environments most relevant to our study. In a scenario reminiscent of bow shocks \citep[][also illustrated by middle part of Figure \ref{fig:ysowinds}]{Smith2003}, the point of impact between the stellar wind and outflow cavity walls will be dissociative and the shock will be J-type. 
The stellar wind colliding with the envelope material also becomes deflected by the cavity walls and sweeps up the surface layers in an entrained flow. The emission from the entrained flow is what is traced by different velocity components of the observed spectra, and these components are what we refer to for the remainder of the paper when discussing velocities.
In the absence of other dissociating sources such as UV fields the J-shock typically takes place at velocities of $\sim$ 25$-$30 \kms\, \citep[][]{Flower2003,Lesaffre2013}. Beyond this we see no line emission in any of the sources included in this study. Further removed from the impact point the post-shock gas has had time to cool down and molecules have had a chance to reform, while at the same time fresh molecular material from the envelope is being exposed to lower-velocity (secondary) shocks originating in the bow wings of the original shock. Depending on the local magnetic field geometry, these shocks may be C-type or non-dissociative J-type (C-J-type) shocks. 
In our observations this corresponds to line emission at velocities greater than a few km\,s$^{-1}$: the line wings at the lowest velocities ($\sim$ 3$-$15 \kms) probe the entrained envelope material as well as secondary shocks and at higher velocities ($\gtrsim$ 15 \kms) they probe the directly shocked material. The lowest-velocity material ($\lesssim$ 3 \kms) is only related to the quiescent envelope, and is ignored in our analysis.

The physical conditions prevailing in the shocked regions and entrained flows enables a number of chemical and physical processes to happen, two of which are especially relevant to our study and are illustrated in the right part of Figure \ref{fig:ysowinds}.
The sputtering effects (a) are relevant in entrained material where the gas and dust is not utterly destroyed in the process (C-J and C-shocks). Water and methanol ice that is sputtered from dust grains is expected to be present in the gas phase in both C-J- and C-shocked regions i.e. at velocities below $\sim$25 \kms. The gas-phase formation reaction of water (b) is expected to activate when sufficiently high temperatures are reached in the shocked regions, and these regions will have an excess of water compared to solid-phase abundances. This temperature threshold is met within the velocity range occupied by the C- and C-J-shocked regions and any velocity above the threshold region will potentially have water formation happening up until the velocities where the physical conditions become dissociative.

\begin{table*}
\begin{center}
\caption{Approximate observed velocity ranges for the different physical and chemical processes.}
\begin{tabular}{l l l}
\hline
\hline
Approx. $v$ range	& Physical process
					& Chemical process		\\
\hline
$\sim$ 3$-$10 \kms	& Secondary CJ and C shocks
					& CH$_3$OH and H$_2$O sputtered and partly destroyed during sputtering \\
					&
					& CH$_3$OH destruction $>$ H$_2$O destruction \\
$\sim$ 10$-$25 \kms & Secondary CJ and C shocks
					& CH$_3$OH sputtered + partly destroyed during sputtering/through reactions with H \\
					&
					& H$_2$O sputtered, partly destroyed by sputtering but also reformed \\
$\sim$ 3$-$10 \kms	& Secondary shocks entraining envelope material
					& Additional sputtering and associated destruction? \\
$\gtrsim$ 25 \kms	& Dissociative shocks
					& CH$_3$OH and H$_2$O destroyed completely; \\
					&
					& only H$_2$O reforms and is offset in velocity by the shock \\
\hline
\label{tab:velstruct}
\end{tabular}
\end{center}
\end{table*}

This translation of velocity components in the spectra into different physical regions in the shock outflow cavity (Summarised in Table \ref{tab:velstruct}) provides us with a sufficient context for the rest of discussion, in which we look at the results of the previous sections to answer the two questions we asked at the beginning of this paper:
\begin{enumerate}
\item How much, if any, methanol is being destroyed in the outflows?
\item Is high-temperature water formation taking place in the outflows?
\end{enumerate}

\subsection{Molecular destruction in outflows}
\label{sec:destruction}
In dissociative J-type shocks, reactions with H are the most efficient way to destroy CH$_3$OH and
other grain products such as water and ammonia. The activation energy for methanol reacting with H ($\sim$ 2270$-$3070 K, \citealt{Sander2011}) is even lower than that of ammonia 
which has already been shown to be destroyed in post-shock gas \citep{Viti2011}. Destroying water in a similar manner is more inefficient because the
activation energy is higher ($\sim$ 10$^4$ K), but at shock velocities of 25$-$30 \kms\ the kinetic
temperature is already $>$ 2$\times$10$^4$ K \citep{McKee1980} and so H$_2$O will also collisionally dissociate.
We can readily observe this difference in activation energies between CH$_3$OH and H$_2$O
destruction by H (Fig. \ref{fig:decomp_h2o}). We see that in all cases
H$_2$O emission in the wings tapers off below the $1-\sigma$ noise level at significantly higher velocities
($\Delta v_{\rm LSR}> 25\ {\rm km\, s^{-1}}$) than the CH$_3$OH emission, which drops below noise levels at approximately $\Delta v_{\rm LSR}=10\ {\rm km\,s^{-1}}$ in most cases. The exception to
this is IRAS 4A, which still shows trace methanol emission at up to $18\ {\rm km\,s^{-1}}$ and
water emission at approximately $35\ {\rm km\,s^{-1}}$.

The gas-phase CH$_3$OH abundances derived from the middle row of Figure \ref{fig:radex_convert} can be compared to the expected solid phase CH$_3$OH abundances. If we assume that equal abundances of envelope CH$_3$OH originate from dust grains at both the low-velocity and high-velocity regimes, the comparison leads to the conclusion that, towards high velocities, some fraction of CH$_3$OH originating from grain surfaces (90$-$99\%) is missing in the gas phase. This CH$_3$OH is either not being sputtered from the grains, or is being destroyed in the sputtering process or in the gas phase.

Sputtering occurs when a neutral species, typically H$_2$, H or He, collides with the ice mantle of a
dust grain with enough kinetic energy to release the ice species into the gas phase. Sputtering at
shock velocities down to 10 \kms\ predominantly takes place in C-type shocks \citep{Flower2010},
where the negatively charged dust grains stream past the neutral gas. At these velocities the kinetic
energy of each particle is of the order of 1 eV. Sputtering yields have been calculated theoretically,
but not measured in the laboratory, although 1 eV is significantly higher than the desorption energy barrier of CH$_3$OH or H$_2$O (both below 0.5 eV, \citealt{Burke2010, Fraser2001}). However, given the high kinetic energy, it is likely that at least
some molecules could be destroyed, rather than simply being desorbed by the sputtering process. This is more likely for CH$_3$OH than H$_2$O, given that the gas-phase dissociation energy of CH$_3$OH to H+ CH$_2$OH is around 4 eV compared to just over 5 eV for H + OH \citep{Blanksby2003}. Furthermore, strong evidence exists that dissociation barriers are lowered in the condensed phase \citep{Woon2002}. However, these sputtering processes are occurring in the same regions where we also expect methanol destruction by high-temperature gas phase reactions with hydrogen atoms. Consequently, although our observations cannot distinguish between the two processes, it is is clear that as the methanol abundance drops in the line wings, the sputtered methanol is being destroyed more readily than the water.

\subsection{Water formation in outflows}
\label{sec:waterformation}

At the highest outflow velocities, higher than those where we observe any methanol emission, water is readily observed. As discussed in Section \ref{sec:destruction}, any water at even higher velocities will be destroyed by the J-shock and this is suggested by the lack of detectable H$_2$O emission beyond $\sim$ 25$-$30 \kms. Though the water destroyed in the J-shock is expected to eventually re-form via the high-temperature gas-phase route, the velocity of the re-formed water has been shifted to only be observable in the ``offset" component directly tracing the shock itself. This component has been removed in the decomposition of the H$_2$O spectra used in this analysis, and has been separately discussed in \cite{Kristensen2013}.
The water at below J-shock velocities must have formed through gas-phase synthesis directly in the warm shocked gas. In our data the column density ratio of H$_2$O/CO starts increasing at velocities higher than $\sim$ 10 \kms\, with respect to the source velocity. We conclude that the shift between water release from the grains and gas-phase water synthesis switches on at this velocity, since the water abundance remains constant with respect to CO \citep{Yldz2013} and CO is not destroyed unless the shocks have significantly higher velocities \citep[$\sim$80 \kms ;][]{Neufeld1989}. This is especially apparent in the blue wing of IRAS 4A and the red wing of IRAS 4B; the water abundance increases by a factor of $\sim$2, due to gas phase formation.
Below $\sim$10 \kms our data shows that the water abundance remains constant with respect to CO and we interpret this to indicate gas-phase water (and methanol) at low velocities which originated entirely from sputtered ice. There are two possibilities for this low-velocity component: (i) if we consider the outflows to be conical shells \citep[e.g.][]{Cabrit1986,Cabrit1990} then a part of the cone, when projected onto the plane of the sky, will have a radial velocity component close to zero even if the 3D velocity in the cone is not zero anywhere; (ii) the post-shock or entrained material has decelerated and cooled down but has not had time to freeze out yet \citep{Bergin1998}. A combination of both possibilities may naturally also be at play. Observations at higher angular resolution are required to break this degeneracy by pinpointing where the emission from the lower-velocity line wings originate spatially: inside the outflow cavity (option 1) or closer to the envelope (option 2).

\section{Conclusions}
\label{sec:conclusions}

We combined observations of H$_2$O, CH$_3$OH and CO
line emission with
\textsc{RADEX} simulations to constrain the gas-phase
abundance variations of water and methanol towards
three low-mass YSOs: IRAS 2A, IRAS 4A and IRAS 4B. We find that:

\begin{enumerate}

\item The CH$_3$OH/CO column density ratio decreases by up to one order of magnitude with increasing velocity.
\item The abundance of CH$_3$OH in the shocked gas is more than 90$\%$ lower than that reported for CH$_3$OH ice in the cold envelope. Given that the CO abundance is unaffected by shock chemistry across the velocity range investigated, this implies that CH$_3$OH is destroyed either during the ice sputtering process from dust grains or through gas-phase reactions with H in the outflow.
\item The H$_2$O/CO column density ratios increase by $\sim$ a factor of two, with the increase beginning at velocities above $\sim$10 \kms. This suggests that the gas-phase formation of water is active and significant at velocities higher than 10 \kms. No discontinuity is observed in the water column density as a function of velocity, implying that the transition is smooth and continuous between the observed water originating from ice sputtering and gas-phase routes.
\item The column density ratio of CH$_3$OH/H$_2$O also decreases with increasing velocity, closely matching the same trend in CH$_3$OH/CO column density ratio. Based on our prior conclusions, this implies that both H$_2$O and CH$_3$OH are sputtered from ices in these shocks, but that only CH$_3$OH is being destroyed, and no gas-phase H$_2$O formation is occurring, at least at velocities below 10 \kms.
\item In the blue wing of IRAS 4A, the CH$_3$OH/H$_2$O column density ratio decreases more steeply at velocities in excess of $\sim$10 \kms\, than the CH$_3$OH/CO column density ratio. This traces the high-temperature gas-phase formation of water in this higher velocity regime.

\end{enumerate}

Consequently, gas-phase CH$_3$OH and H$_2$O abundances in shocked regions depend on the complex interplay between ice sputtering mechanisms, gas-phase destruction processes and high-temperature formation reactions.
These conclusions hint towards future
observational and experimental requirements to further constrain the physics and chemistry of H$_{2}$O and CH$_{3}$OH in outflows.
In particular, trying to observe and quantify the efficiency of CH$_3$OH and H$_2$O ice sputtering
by neutrals in a laboratory environment would
contribute towards verifying and/or refuting two of our
main conclusions; whether or not the two molecules are sputtered with equal efficiency, and if CH$_3$OH can be destroyed in the sputtering processes. Observations resolving the outflow spatially close to the protostar, and mapping the variation of physical conditions, are also important to further constrain these results.

\section*{Acknowledgements}
ANS acknowledges financial support by the
European Community FP7-ITN Marie-Curie Programme
(grant agreement 238258). Astrochemistry in Leiden is supported by
the Netherlands Research School for Astronomy (NOVA), by a Spinoza grant
and by the European Community's Seventh Framework Programme FP7/2007-
2013 under grant agreement 238258 (LASSIE). HIFI has been designed and
built by a consortium of institutes and university departments from across Europe,
Canada and the United States under the leadership of SRON Netherlands
Institute for Space Research, Groningen, The Netherlands and with major contributions
from Germany, France and the US. Consortium members are: Canada:
CSA, U.Waterloo; France: CESR, LAB, LERMA, IRAM; Germany: KOSMA,
MPIfR, MPS; Ireland, NUI Maynooth; Italy: ASI, IFSI-INAF, Osservatorio Astrofisico
di Arcetri- INAF; Netherlands: SRON, TUD; Poland: CAMK, CBK;
Spain: Observatorio Astronomico Nacional (IGN), Centro de Astrobiologia
(CSIC-INTA). Sweden: Chalmers University of Technology - MC2, RSS \&
GARD; Onsala Space Observatory; Swedish National Space Board, Stockholm
University - Stockholm Observatory; Switzerland: ETH Zurich, FHNW; USA:
Caltech, JPL, NHSC. The authors thank the referee for their helpful comments and questions.

\bibliographystyle{mn2e}

\bibliography{ysocomp_references}

\begin{thebibliography}{75}
\expandafter\ifx\csname natexlab\endcsname\relax\def\natexlab#1{#1}\fi

\bibitem[{Bachiller {et~al}\mbox{.}(1995)Bachiller, Liechti, Walmsley, \&
  Colomer}]{Bachiller1995}
Bachiller R., Liechti S., Walmsley C.~M., Colomer F., 1995, Astronomy \&
  Astrophysics, 295, L51

\bibitem[{{Bergin}, {Neufeld} \& {Melnick}(1998){Bergin}, {Neufeld}, \&
  {Melnick}}]{Bergin1998}
{Bergin} E.~A., {Neufeld} D.~A., {Melnick} G.~J., 1998, The Astrophysical
  Journal, 499, 777

\bibitem[{Bergin, Neufeld \& Melnick(1999)Bergin, Neufeld, \&
  Melnick}]{Bergin1999}
Bergin E.~A., Neufeld D.~A., Melnick G.~J., 1999, The Astrophysical Journal,
  510, L145

\bibitem[{{Bjerkeli} {et~al}\mbox{.}(2012){Bjerkeli}, {Liseau}, {Larsson},
  {Rydbeck}, {Nisini}, {Tafalla}, {Antoniucci}, {Benedettini}, {Bergman},
  {Cabrit}, {Giannini}, {Melnick}, {Neufeld}, {Santangelo}, \& {van
  Dishoeck}}]{Bjerkeli2012}
{Bjerkeli} P. {et~al.}, 2012, Astronomy \& Astrophysics, 546, A29

\bibitem[{Blanksby \& Ellison(2003)}]{Blanksby2003}
Blanksby S.~J., Ellison G.~B., 2003, Accounts of Chemical Research, 36, 25

\bibitem[{Boogert {et~al}\mbox{.}(2008)Boogert, Pontoppidan, Knez, Lahuis,
  Kessler‐Silacci, van Dishoeck, Blake, Augereau, Bisschop, Bottinelli,
  Brooke, Brown, Crapsi, {Evans II}, Fraser, Geers, Huard, J\o~rgensen,
  \"{O}berg, Allen, Harvey, Koerner, Mundy, Padgett, Sargent, \&
  Stapelfeldt}]{Boogert2008}
Boogert A. C.~A. {et~al.}, 2008, The Astrophysical Journal, 678, 985

\bibitem[{Brown \& Bolina(2007)}]{Brown2007}
Brown W.~A., Bolina A.~S., 2007, Monthly Notices of the Royal Astronomical
  Society, 374, 1006

\bibitem[{Burke \& Brown(2010)}]{Burke2010}
Burke D.~J., Brown W.~A., 2010, Physical Chemistry Chemical Physics, 12, 5947

\bibitem[{{Cabrit} \& {Bertout}(1986)}]{Cabrit1986}
{Cabrit} S., {Bertout} C., 1986, The Astrophysical Journal, 307, 313

\bibitem[{{Cabrit} \& {Bertout}(1990)}]{Cabrit1990}
{Cabrit} S., {Bertout} C., 1990, The Astrophysical Journal, 348, 530

\bibitem[{Caselli {et~al}\mbox{.}(2012)Caselli, Keto, Bergin, Tafalla, Aikawa,
  Douglas, Pagani, Y\'{\i}ld\'{\i}z, van~der Tak, Walmsley, Codella, Nisini,
  Kristensen, \& van Dishoeck}]{Caselli2012}
Caselli P. {et~al.}, 2012, The Astrophysical Journal, 759, L37

\bibitem[{{Charnley}(1999)}]{Charnley1999}
{Charnley} S.~B., 1999, in NATO ASIC Proc. 523: Formation and Evolution of
  Solids in Space, p. 131

\bibitem[{Codella {et~al}\mbox{.}(2010)Codella, Lefloch, Ceccarelli,
  Cernicharo, Caux, Lorenzani, Viti, Hily-Blant, Parise, Maret, Nisini,
  Caselli, Cabrit, Pagani, Benedettini, Boogert, Gueth, Melnick, Neufeld,
  Pacheco, Salez, Schuster, Bacmann, Baudry, Bell, Bergin, Blake, Bottinelli,
  Castets, Comito, Coutens, Crimier, Dominik, Demyk, Encrenaz, Falgarone,
  Fuente, Gerin, Goldsmith, Helmich, Hennebelle, Henning, Herbst, Jacq, Kahane,
  Kama, Klotz, Langer, Lis, Lord, Pearson, Phillips, Saraceno, Schilke,
  Tielens, van~der Tak, van~der Wiel, Vastel, Wakelam, Walters, Wyrowski,
  Yorke, Borys, Delorme, Kramer, Larsson, Mehdi, Ossenkopf, \&
  Stutzki}]{Codella2010}
Codella C. {et~al.}, 2010, Astronomy \& Astrophysics, 518, L112

\bibitem[{Cuppen {et~al}\mbox{.}(2009)Cuppen, van Dishoeck, Herbst, \&
  Tielens}]{Cuppen2009}
Cuppen H.~M., van Dishoeck E.~F., Herbst E., Tielens A. G. G.~M., 2009,
  Astronomy \& Astrophysics, 508, 275

\bibitem[{Daniel, Dubernet \& Grosjean(2011)Daniel, Dubernet, \&
  Grosjean}]{Daniel2011}
Daniel F., Dubernet M.-L., Grosjean A., 2011, Astronomy \& Astrophysics, 536,
  A76

\bibitem[{Dartois {et~al}\mbox{.}(1999)Dartois, Schutte, Geballe, Demyk,
  Ehrenfreund, \& D'Hendecourt}]{Dartois1999}
Dartois E., Schutte W., Geballe T.~R., Demyk K., Ehrenfreund P., D'Hendecourt
  L., 1999, Astronomy \& Astrophysics, 342, L32

\bibitem[{de~Graauw {et~al}\mbox{.}(2010)de~Graauw, Helmich, Phillips, Stutzki,
  Caux, Whyborn, Dieleman, Roelfsema, Aarts, Assendorp, Bachiller, Baechtold,
  Barcia, Beintema, Belitsky, Benz, Bieber, Boogert, Borys, Bumble, Ca\"{\i}s,
  Caris, Cerulli-Irelli, Chattopadhyay, Cherednichenko, Ciechanowicz,
  Coeur-Joly, Comito, Cros, de~Jonge, de~Lange, Delforges, Delorme, den
  Boggende, Desbat, Diez-Gonz\'{a}lez, {Di Giorgio}, Dubbeldam, Edwards,
  Eggens, Erickson, Evers, Fich, Finn, Franke, Gaier, Gal, Gao, Gallego,
  Gauffre, Gill, Glenz, Golstein, Goulooze, Gunsing, G\"{u}sten, Hartogh,
  Hatch, Higgins, Honingh, Huisman, Jackson, Jacobs, Jacobs, Jarchow, Javadi,
  Jellema, Justen, Karpov, Kasemann, Kawamura, Keizer, Kester, Klapwijk, Klein,
  Kollberg, Kooi, Kooiman, Kopf, Krause, Krieg, Kramer, Kruizenga, Kuhn,
  Laauwen, Lai, Larsson, Leduc, Leinz, Lin, Liseau, Liu, Loose,
  L\'{o}pez-Fernandez, Lord, Luinge, Marston, Mart\'{\i}n-Pintado, Maestrini,
  Maiwald, McCoey, Mehdi, Megej, Melchior, Meinsma, Merkel, Michalska,
  Monstein, Moratschke, Morris, Muller, Murphy, Naber, Natale, Nowosielski,
  Nuzzolo, Olberg, Olbrich, Orfei, Orleanski, Ossenkopf, Peacock, Pearson,
  Peron, Phillip-May, Piazzo, Planesas, Rataj, Ravera, Risacher, Salez,
  Samoska, Saraceno, Schieder, Schlecht, Schl\"{o}der, Schm\"{u}lling, Schultz,
  Schuster, Siebertz, Smit, Szczerba, Shipman, Steinmetz, Stern, Stokroos,
  Teipen, Teyssier, Tils, Trappe, van Baaren, van Leeuwen, van~de Stadt,
  Visser, Wildeman, Wafelbakker, Ward, Wesselius, Wild, Wulff, Wunsch, Tielens,
  Zaal, Zirath, Zmuidzinas, \& Zwart}]{deGraauw2010}
de~Graauw T. {et~al.}, 2010, Astronomy \& Astrophysics, 518, L6

\bibitem[{{Dionatos} {et~al}\mbox{.}(2013){Dionatos}, {J{\o}rgensen}, {Green},
  {Herczeg}, {Evans}, {Kristensen}, {Lindberg}, \& {van
  Dishoeck}}]{Dionatos2013}
{Dionatos} O., {J{\o}rgensen} J.~K., {Green} J.~D., {Herczeg} G.~J., {Evans}
  N.~J., {Kristensen} L.~E., {Lindberg} J.~E., {van Dishoeck} E.~F., 2013,
  Astronomy \& Astrophysics, 558, A88

\bibitem[{Draine, Roberge \& Dalgarno(1983)Draine, Roberge, \&
  Dalgarno}]{Draine1983}
Draine B.~T., Roberge W.~G., Dalgarno A., 1983, The Astrophysical Journal, 264,
  485

\bibitem[{{Emprechtinger} {et~al}\mbox{.}(2013){Emprechtinger}, {Lis},
  {Rolffs}, {Schilke}, {Monje}, {Comito}, {Ceccarelli}, {Neufeld}, \& {van der
  Tak}}]{Emprechtinger2013}
{Emprechtinger} M. {et~al.}, 2013, The Astrophysical Journal, 765, 61

\bibitem[{{Flower} {et~al}\mbox{.}(2003){Flower}, {Le Bourlot}, {Pineau des
  For{\^e}ts}, \& {Cabrit}}]{Flower2003}
{Flower} D.~R., {Le Bourlot} J., {Pineau des For{\^e}ts} G., {Cabrit} S., 2003,
  Monthly Notices of the Royal Astronomical Society, 341, 70

\bibitem[{Flower \& {Pineau des For\^{e}ts}(2010)}]{Flower2010}
Flower D.~R., {Pineau des For\^{e}ts} G., 2010, Monthly Notices of the Royal
  Astronomical Society, 406, 1745

\bibitem[{Fraser {et~al}\mbox{.}(2001)Fraser, Collings, McCoustra, \&
  Williams}]{Fraser2001}
Fraser H.~J., Collings M.~P., McCoustra M. R.~S., Williams D.~A., 2001, Monthly
  Notices of the Royal Astronomical Society, 327, 1165

\bibitem[{Garrod {et~al}\mbox{.}(2006)Garrod, {Hee Park}, Caselli, \&
  Herbst}]{Garrod2006}
Garrod R., {Hee Park} I., Caselli P., Herbst E., 2006, Faraday Discussions,
  133, 51

\bibitem[{Geppert {et~al}\mbox{.}(2006)Geppert, Hamberg, Thomas, \"{O}sterdahl,
  Hellberg, Zhaunerchyk, Ehlerding, Millar, Roberts, Semaniak, af~Ugglas,
  K\"{a}llberg, Simonsson, Kaminska, \& Larsson}]{Geppert2006}
Geppert W.~D. {et~al.}, 2006, Faraday Discussions, 133, 177

\bibitem[{Gibb {et~al}\mbox{.}(2004)Gibb, Whittet, Boogert, \&
  Tielens}]{Gibb2004}
Gibb E.~L., Whittet D. C.~B., Boogert A. C.~A., Tielens A. G. G.~M., 2004, The
  Astrophysical Journal Supplement Series, 151, 35

\bibitem[{Glassgold, Meijerink \& Najita(2009)Glassgold, Meijerink, \&
  Najita}]{Glassgold2009}
Glassgold A.~E., Meijerink R., Najita J.~R., 2009, The Astrophysical Journal,
  701, 142

\bibitem[{{Green} {et~al}\mbox{.}(2013){Green}, {Evans}, {J{\o}rgensen},
  {Herczeg}, {Kristensen}, {Lee}, {Dionatos}, {Yildiz}, {Salyk}, {Meeus},
  {Bouwman}, {Visser}, {Bergin}, {van Dishoeck}, {Rascati}, {Karska}, {van
  Kempen}, {Dunham}, {Lindberg}, {Fedele}, \& {DIGIT Team}}]{Green2013}
{Green} J.~D. {et~al.}, 2013, The Astrophysical Journal, 770, 123

\bibitem[{{Herczeg} {et~al}\mbox{.}(2012){Herczeg}, {Karska}, {Bruderer},
  {Kristensen}, {van Dishoeck}, {J{\o}rgensen}, {Visser}, {Wampfler}, {Bergin},
  {Y{\i}ld{\i}z}, {Pontoppidan}, \& {Gracia-Carpio}}]{Herzceg2012}
{Herczeg} G.~J. {et~al.}, 2012, Astronomy \& Astrophysics, 540, A84

\bibitem[{{Herpin} {et~al}\mbox{.}(2012){Herpin}, {Chavarr{\'{\i}}a}, {van der
  Tak}, {Wyrowski}, {van Dishoeck}, {Jacq}, {Braine}, {Baudry}, {Bontemps}, \&
  {Kristensen}}]{Herpin2012}
{Herpin} F. {et~al.}, 2012, Astronomy \& Astrophysics, 542, A76

\bibitem[{Ioppolo {et~al}\mbox{.}(2008)Ioppolo, Cuppen, Romanzin, van Dishoeck,
  \& Linnartz}]{Ioppolo2008}
Ioppolo S., Cuppen H.~M., Romanzin C., van Dishoeck E.~F., Linnartz H., 2008,
  The Astrophysical Journal, 686, 1474

\bibitem[{Karska {et~al}\mbox{.}(2013)Karska, Herczeg, van Dishoeck, Wampfler,
  Kristensen, Goicoechea, Visser, Nisini, {San Jos\'{e}-Garc\'{\i}a}, Bruderer,
  \'{S}niady, Doty, Fedele, Yıldız, Benz, Bergin, Caselli, Herpin,
  Hogerheijde, Johnstone, J\o~rgensen, Liseau, Tafalla, van~der Tak, \&
  Wyrowski}]{Karska2013}
Karska A. {et~al.}, 2013, Astronomy \& Astrophysics, 552, A141

\bibitem[{Kristensen {et~al}\mbox{.}(2013)Kristensen, van Dishoeck, Benz,
  Bruderer, Visser, \& Wampfler}]{Kristensen2013}
Kristensen L.~E., van Dishoeck E.~F., Benz A.~O., Bruderer S., Visser R.,
  Wampfler S.~F., 2013, Astronomy \& Astrophysics, 557, A23

\bibitem[{Kristensen {et~al}\mbox{.}(2012)Kristensen, van Dishoeck, Bergin,
  Visser, Yıldız, {San Jose-Garcia}, J\o~rgensen, Herczeg, Johnstone,
  Wampfler, Benz, Bruderer, Cabrit, Caselli, Doty, Harsono, Herpin,
  Hogerheijde, Karska, van Kempen, Liseau, Nisini, Tafalla, van~der Tak, \&
  Wyrowski}]{Kristensen2012}
Kristensen L.~E. {et~al.}, 2012, Astronomy \& Astrophysics, 542, A8

\bibitem[{{Kristensen} {et~al}\mbox{.}(2011){Kristensen}, {van Dishoeck},
  {Tafalla}, {Bachiller}, {Nisini}, {Liseau}, \&
  {Y{\i}ld{\i}z}}]{Kristensen2011}
{Kristensen} L.~E., {van Dishoeck} E.~F., {Tafalla} M., {Bachiller} R.,
  {Nisini} B., {Liseau} R., {Y{\i}ld{\i}z} U.~A., 2011, Astronomy \&
  Astrophysics, 531, L1

\bibitem[{Kristensen {et~al}\mbox{.}(2010{\natexlab{a}})Kristensen, van
  Dishoeck, van Kempen, Cuppen, Brinch, J\o~rgensen, \&
  Hogerheijde}]{Kristensen2010}
Kristensen L.~E., van Dishoeck E.~F., van Kempen T.~A., Cuppen H.~M., Brinch
  C., J\o~rgensen J.~K., Hogerheijde M.~R., 2010{\natexlab{a}}, Astronomy \&
  Astrophysics, 516, A57

\bibitem[{Kristensen {et~al}\mbox{.}(2010{\natexlab{b}})Kristensen, Visser, van
  Dishoeck, Yıldız, Doty, Herczeg, Liu, Parise, J\o~rgensen, van Kempen,
  Brinch, Wampfler, Bruderer, Benz, Hogerheijde, Deul, Bachiller, Baudry,
  Benedettini, Bergin, Bjerkeli, Blake, Bontemps, Braine, Caselli, Cernicharo,
  Codella, Daniel, de~Graauw, di~Giorgio, Dominik, Encrenaz, Fich, Fuente,
  Giannini, Goicoechea, Helmich, Herpin, Jacq, Johnstone, Kaufman, Larsson,
  Lis, Liseau, Marseille, McCoey, Melnick, Neufeld, Nisini, Olberg, Pearson,
  Plume, Risacher, Santiago-Garc\'{\i}a, Saraceno, Shipman, Tafalla, Tielens,
  van~der Tak, Wyrowski, Beintema, de~Jonge, Dieleman, Ossenkopf, Roelfsema,
  Stutzki, \& Whyborn}]{Kristensen2010b}
Kristensen L.~E. {et~al.}, 2010{\natexlab{b}}, Astronomy \& Astrophysics, 521,
  L30

\bibitem[{{Lacy} {et~al}\mbox{.}(1994){Lacy}, {Knacke}, {Geballe}, \&
  {Tokunaga}}]{Lacy1994}
{Lacy} J.~H., {Knacke} R., {Geballe} T.~R., {Tokunaga} A.~T., 1994, The
  Astrophysical Journal, 428, L69

\bibitem[{Lamberts {et~al}\mbox{.}(2013)Lamberts, Cuppen, Ioppolo, \&
  Linnartz}]{Lamberts2013}
Lamberts T., Cuppen H.~M., Ioppolo S., Linnartz H., 2013, Physical Chemistry
  Chemical Physics, 15, 8287

\bibitem[{Lee, Bettens \& Herbst(1996)Lee, Bettens, \& Herbst}]{LeeH.-H.1996}
Lee H.-H., Bettens R. P.~A., Herbst E., 1996, Astronomy \& Astrophysics
  Supplement, 119, 111

\bibitem[{{Lesaffre} {et~al}\mbox{.}(2013){Lesaffre}, {Pineau des For{\^e}ts},
  {Godard}, {Guillard}, {Boulanger}, \& {Falgarone}}]{Lesaffre2013}
{Lesaffre} P., {Pineau des For{\^e}ts} G., {Godard} B., {Guillard} P.,
  {Boulanger} F., {Falgarone} E., 2013, Astronomy \& Astrophysics, 550, A106

\bibitem[{{Manoj} {et~al}\mbox{.}(2013){Manoj}, {Watson}, {Neufeld}, {Megeath},
  {Vavrek}, {Yu}, {Visser}, {Bergin}, {Fischer}, {Tobin}, {Stutz}, {Ali},
  {Wilson}, {Di Francesco}, {Osorio}, {Maret}, \& {Poteet}}]{Manoj2013}
{Manoj} P. {et~al.}, 2013, The Astrophysical Journal, 763, 83

\bibitem[{Maret {et~al}\mbox{.}(2005)Maret, Ceccarelli, Tielens, Caux, Lefloch,
  Faure, Castets, \& Flower}]{Maret2005}
Maret S., Ceccarelli C., Tielens A. G. G.~M., Caux E., Lefloch B., Faure A.,
  Castets A., Flower D.~R., 2005, Astronomy \& Astrophysics, 442, 527

\bibitem[{{McKee} \& {Hollenbach}(1980)}]{McKee1980}
{McKee} C.~F., {Hollenbach} D.~J., 1980, Annual Review of Astronomy \&
  Astrophysics, 18, 219

\bibitem[{{Mottram} {et~al}\mbox{.}(2013){Mottram}, {van Dishoeck}, {Schmalzl},
  {Kristensen}, {Visser}, {Hogerheijde}, \& {Bruderer}}]{Mottram2013}
{Mottram} J.~C., {van Dishoeck} E.~F., {Schmalzl} M., {Kristensen} L.~E.,
  {Visser} R., {Hogerheijde} M.~R., {Bruderer} S., 2013, Astronomy \&
  Astrophysics, 558, A126

\bibitem[{{Neufeld} \& {Dalgarno}(1989)}]{Neufeld1989}
{Neufeld} D.~A., {Dalgarno} A., 1989, The Astrophysical Journal, 340, 869

\bibitem[{{Nisini} {et~al}\mbox{.}(2010){Nisini}, {Benedettini}, {Codella},
  {Giannini}, {Liseau}, {Neufeld}, {Tafalla}, {van Dishoeck}, {Bachiller},
  {Baudry}, {Benz}, {Bergin}, {Bjerkeli}, {Blake}, {Bontemps}, {Braine},
  {Bruderer}, {Caselli}, {Cernicharo}, {Daniel}, {Encrenaz}, {di Giorgio},
  {Dominik}, {Doty}, {Fich}, {Fuente}, {Goicoechea}, {de Graauw}, {Helmich},
  {Herczeg}, {Herpin}, {Hogerheijde}, {Jacq}, {Johnstone}, {J{\o}rgensen},
  {Kaufman}, {Kristensen}, {Larsson}, {Lis}, {Marseille}, {McCoey}, {Melnick},
  {Olberg}, {Parise}, {Pearson}, {Plume}, {Risacher}, {Santiago}, {Saraceno},
  {Shipman}, {van Kempen}, {Visser}, {Viti}, {Wampfler}, {Wyrowski}, {van der
  Tak}, {Y{\i}ld{\i}z}, {Delforge}, {Desbat}, {Hatch}, {P{\'e}ron}, {Schieder},
  {Stern}, {Teyssier}, \& {Whyborn}}]{Nisini2010}
{Nisini} B. {et~al.}, 2010, Astronomy \& Astrophysics, 518, L120

\bibitem[{{Nisini} {et~al}\mbox{.}(2013){Nisini}, {Santangelo}, {Antoniucci},
  {Benedettini}, {Codella}, {Giannini}, {Lorenzani}, {Liseau}, {Tafalla},
  {Bjerkeli}, {Cabrit}, {Caselli}, {Kristensen}, {Neufeld}, {Melnick}, \& {van
  Dishoeck}}]{Nisini2013}
{Nisini} B. {et~al.}, 2013, Astronomy \& Astrophysics, 549, A16

\bibitem[{\"{O}berg {et~al}\mbox{.}(2011)\"{O}berg, Boogert, Pontoppidan,
  van~den Broek, van Dishoeck, Bottinelli, Blake, \& Evans}]{Oberg2011}
\"{O}berg K.~I., Boogert A. C.~A., Pontoppidan K.~M., van~den Broek S., van
  Dishoeck E.~F., Bottinelli S., Blake G.~A., Evans N.~J., 2011, The
  Astrophysical Journal, 740, 109

\bibitem[{Pickett {et~al}\mbox{.}(1998)Pickett, Poynter, Cohen, Delitsky,
  Pearson, \& Muller}]{JPLcat}
Pickett H.~M., Poynter R.~L., Cohen E.~A., Delitsky M.~L., Pearson J.~C.,
  Muller H. S.~P., 1998, J. Quant. Spectrosc. \& Rad. Transfer, 60, 883

\bibitem[{Pilbratt {et~al}\mbox{.}(2010)Pilbratt, Riedinger, Passvogel, Crone,
  Doyle, Gageur, Heras, Jewell, Metcalfe, Ott, \& Schmidt}]{Pilbratt2010}
Pilbratt G.~L. {et~al.}, 2010, Astronomy \& Astrophysics, 518, L1

\bibitem[{Pontoppidan {et~al}\mbox{.}(2003)Pontoppidan, Dartois, van Dishoeck,
  Thi, \& D'Hendecourt}]{Pontoppidan2003}
Pontoppidan K.~M., Dartois E., van Dishoeck E.~F., Thi W.-F., D'Hendecourt L.,
  2003, Astronomy \& Astrophysics, 404, L17

\bibitem[{Pontoppidan, van Dishoeck \& Dartois(2004)Pontoppidan, van Dishoeck,
  \& Dartois}]{Pontoppidan2004}
Pontoppidan K.~M., van Dishoeck E.~F., Dartois E., 2004, Astronomy \&
  Astrophysics, 426, 925

\bibitem[{Rabli \& Flower(2010)}]{Rabli2010}
Rabli D., Flower D.~R., 2010, Monthly Notices of the Royal Astronomical
  Society, 406, 95

\bibitem[{Roelfsema {et~al}\mbox{.}(2011)Roelfsema, Helmich, Teyssier,
  Ossenkopf, Morris, Olberg, Shipman, Risacher, Akyilmaz, Assendorp, Avruch,
  Beintema, Biver, Boogert, Borys, Braine, Caris, Caux, Cernicharo, Coeur-Joly,
  Comito, de~Lange, Delforge, Dieleman, Dubbeldam, de~Graauw, Edwards, Fich,
  Flederus, Gal, di~Giorgio, Herpin, Higgins, Hoac, Huisman, Jarchow, Jellema,
  de~Jonge, Kester, Klein, Kooi, Kramer, Laauwen, Larsson, Leinz, Lord,
  Lorenzani, Luinge, Marston, Mart\'{\i}n-Pintado, McCoey, Melchior, Michalska,
  Moreno, M\"{u}ller, Nowosielski, Okada, Orleański, Phillips, Pearson,
  Rabois, Ravera, Rector, Rengel, Sagawa, Salomons, S\'{a}nchez-Su\'{a}rez,
  Schieder, Schl\"{o}der, Schm\"{u}lling, Soldati, Stutzki, Thomas, Tielens,
  Vastel, Wildeman, Xie, Xilouris, Wafelbakker, Whyborn, Zaal, Bell, Bjerkeli,
  de~Beck, Cavali\'{e}, Crockett, Hily-Blant, Kama, Kaminski, Lefl\'{o}ch,
  Lombaert, {De Luca}, Makai, Marseille, Nagy, Pacheco, van~der Wiel, Wang, \&
  Yıldız}]{Roelfsema2012}
Roelfsema P.~R. {et~al.}, 2011, Astronomy \& Astrophysics, 537, A17

\bibitem[{Sander {et~al}\mbox{.}(2011)Sander, Abbatt, Barker, Burkholder,
  Friedl, Golden, Huie, Kolb, Kurylo, Moortgat, Orkin, \& Wine}]{Sander2011}
Sander S. {et~al.}, 2011, JPL Publication, 10, 6

\bibitem[{Santangelo {et~al}\mbox{.}(2013)Santangelo, Nisini, Antoniucci,
  Codella, Cabrit, Giannini, Herczeg, Liseau, Tafalla, \& van
  Dishoeck}]{Santangelo2013}
Santangelo G. {et~al.}, 2013, Astronomy \& Astrophysics, 557, A22

\bibitem[{Santangelo {et~al}\mbox{.}(2012)Santangelo, Nisini, Giannini,
  Antoniucci, Vasta, Codella, Lorenzani, Tafalla, Liseau, van Dishoeck, \&
  Kristensen}]{Santangelo2012}
Santangelo G. {et~al.}, 2012, Astronomy \& Astrophysics, 538, A45

\bibitem[{Sch\"{o}ier {et~al}\mbox{.}(2005)Sch\"{o}ier, van~der Tak, van
  Dishoeck, \& Black}]{Schoier2005}
Sch\"{o}ier F.~L., van~der Tak F. F.~S., van Dishoeck E.~F., Black J.~H., 2005,
  Astronomy \& Astrophysics, 432, 369

\bibitem[{{Smith}, {Khanzadyan} \& {Davis}(2003){Smith}, {Khanzadyan}, \&
  {Davis}}]{Smith2003}
{Smith} M.~D., {Khanzadyan} T., {Davis} C.~J., 2003, Monthly Notices of the
  Royal Astronomical Society, 339, 524

\bibitem[{{Snell} {et~al}\mbox{.}(2000){Snell}, {Howe}, {Ashby}, {Bergin},
  {Chin}, {Erickson}, {Goldsmith}, {Harwit}, {Kleiner}, {Koch}, {Neufeld},
  {Patten}, {Plume}, {Schieder}, {Stauffer}, {Tolls}, {Wang}, {Winnewisser},
  {Zhang}, \& {Melnick}}]{Snell2000}
{Snell} R.~L. {et~al.}, 2000, The Astrophysical Journal, 539, L93

\bibitem[{Tafalla {et~al}\mbox{.}(2013)Tafalla, Liseau, Nisini, Bachiller,
  Santiago-Garc\'{\i}a, van Dishoeck, Kristensen, Herczeg, \&
  Yıldız}]{Tafalla2013}
Tafalla M. {et~al.}, 2013, Astronomy \& Astrophysics, 551, A116

\bibitem[{Tafalla {et~al}\mbox{.}(2010)Tafalla, Santiago-Garcia, Hacar, \&
  Bachiller}]{Tafalla2010}
Tafalla M., Santiago-Garcia J., Hacar A., Bachiller R., 2010, Astronomy \&
  Astrophysics, 522, A91

\bibitem[{Tielens(2005)}]{Tielens2005}
Tielens A. G. G.~M., 2005, {The Physics and Chemistry of the Interstellar
  Medium}. Cambridge University Press

\bibitem[{Tielens \& Hagen(1982)}]{Tielens1982}
Tielens A. G. G.~M., Hagen W., 1982, Astronomy \& Astrophysics, 114, 245

\bibitem[{van~der Tak {et~al}\mbox{.}(2007)van~der Tak, Black, Schoier, Jansen,
  \& van Dishoeck}]{VanderTak2007}
van~der Tak F. F.~S., Black J.~H., Schoier F.~L., Jansen D.~J., van Dishoeck
  E.~F., 2007, Astronomy \& Astrophysics, 468, 627

\bibitem[{van~der Tak, van Dishoeck \& Caselli(2000)van~der Tak, van Dishoeck,
  \& Caselli}]{VanderTak2000}
van~der Tak F. F.~S., van Dishoeck E.~F., Caselli P., 2000, Astronomy \&
  Astrophysics, 361, 327

\bibitem[{van Dishoeck {et~al}\mbox{.}(2011)van Dishoeck, Kristensen, Benz,
  Bergin, Caselli, Cernicharo, Herpin, Hogerheijde, Johnstone, Liseau, Nisini,
  Shipman, Tafalla, van~der Tak, Wyrowski, Aikawa, Bachiller, Baudry,
  Benedettini, Bjerkeli, Blake, Bontemps, Braine, Brinch, Bruderer,
  Chavarr\'{\i}a, Codella, Daniel, de~Graauw, Deul, di~Giorgio, Dominik, Doty,
  Dubernet, Encrenaz, Feuchtgruber, Fich, Frieswijk, Fuente, Giannini,
  Goicoechea, Helmich, Herczeg, Jacq, J\o~rgensen, Karska, Kaufman, Keto,
  Larsson, Lefloch, Lis, Marseille, McCoey, Melnick, Neufeld, Olberg, Pagani,
  Pani\'{c}, Parise, Pearson, Plume, Risacher, Salter, Santiago-Garc\'{\i}a,
  Saraceno, St\"{a}uber, van Kempen, Visser, Viti, Walmsley, Wampfler, \&
  Yıldız}]{vanDishoeck2011}
van Dishoeck E.~F. {et~al.}, 2011, Publications of the Astronomical Society of
  the Pacific, 123, 138

\bibitem[{{Van Loo} {et~al}\mbox{.}(2013){Van Loo}, {Ashmore}, {Caselli},
  {Falle}, \& {Hartquist}}]{VanLoo2013}
{Van Loo} S., {Ashmore} I., {Caselli} P., {Falle} S.~A.~E.~G., {Hartquist}
  T.~W., 2013, Monthly Notices of the Royal Astronomical Society, 428, 381

\bibitem[{Vasta {et~al}\mbox{.}(2012)Vasta, Codella, Lorenzani, Santangelo,
  Nisini, Giannini, Tafalla, Liseau, van Dishoeck, \& Kristensen}]{Vasta2012}
Vasta M. {et~al.}, 2012, Astronomy \& Astrophysics, 537, A98

\bibitem[{{Visser} {et~al}\mbox{.}(2009){Visser}, {van Dishoeck}, {Doty}, \&
  {Dullemond}}]{Visser2009}
{Visser} R., {van Dishoeck} E.~F., {Doty} S.~D., {Dullemond} C.~P., 2009,
  Astronomy \& Astrophysics, 495, 881

\bibitem[{Viti {et~al}\mbox{.}(2011)Viti, Jimenez-Serra, Yates, Codella, Vasta,
  Caselli, Lefloch, \& Ceccarelli}]{Viti2011}
Viti S., Jimenez-Serra I., Yates J.~A., Codella C., Vasta M., Caselli P.,
  Lefloch B., Ceccarelli C., 2011, The Astrophysical Journal, 740, L3

\bibitem[{Woon(2002)}]{Woon2002}
Woon D.~E., 2002, International Journal of Quantum Chemistry, 88, 226

\bibitem[{Yang {et~al}\mbox{.}(2010)Yang, Stancil, Balakrishnan, \&
  Forrey}]{Yang2010}
Yang B., Stancil P.~C., Balakrishnan N., Forrey R.~C., 2010, The Astrophysical
  Journal, 718, 1062

\bibitem[{Y{\i}ld{\i}z {et~al}\mbox{.}(2013)Y{\i}ld{\i}z, Kristensen, van
  Dishoeck, Jose-Garcia, Karska, Harsono, Tafalla, Fuente, Visser, J\o~rgensen,
  \& Hogerheijde}]{Yldz2013}
Y{\i}ld{\i}z U.~A. {et~al.}, 2013, Astronomy \& Astrophysics, 556, 48

\end{thebibliography}

\bsp

\appendix

\section{Correcting for differential beam dilution}
\label{app:beamdilution}
As seen in Table \ref{tab:observations_lines}, the lines that are
examined in this study have somewhat different beamsizes from
each other, ranging from 11\arcsec{} with the CH$_3$OH $5_K-4_K$
lines to 22\arcsec{} with the H$_2$O $2_{02}-1_{11}$ line.
These variations in beamsize lead to large variations from line to
line in the beam dilution of the observed emission, which has
to be compensated for if we are to make an accurate comparison
between the different lines. To do this we have applied to
our line ratios -discussed in Sections \ref{sec:ratios}
and \ref{sec:molvar}- the differential beam dilution correction
method detailed in Appendix B of \citet{Tafalla2010}.

The correction method is applied by assuming a simple
gaussian beam, where the beam dilution factor is proportional
to $\theta_{MB}^{-\alpha}$, where $\theta_{MB}$ is the FWHM
of the beam being examined and $\alpha$ is a free parameter
dependant on the source size and which can vary between 0 and 2.
Tafalla notes that $\alpha=2.0$ for a point-like source $1.0$
for a one-dimensional source and $0.0$ for a source infinitely
extended in two dimensions. For our case, in applying a
beam dilution correction to an observed line intensity ratio
$T_{\rm R,1}/T_{\rm R,2}$ between molecules 1 and 2 can be done
using the equation:
\begin{equation}
(T_{\rm R,1}/T_{\rm R,2})_{\rm corrected} =
(T_{\rm R,1}/T_{\rm R,2})\cdot
(\theta_{\rm MB,2}/\theta_{\rm MB,1})^{-\alpha}.
\end{equation}

In order to apply this correction
method to our data we had to first examine which value of
$\alpha$ is most appropriate to our case.
The effect of accounting for beam dilution, or lack thereof,
in our case is best seen when translating the line ratio of
the two different CH$_3$OH lines against either of the CO lines
or the H$_2$O line we have studied in this paper. For the
case of finding out $\alpha$ we decided to use the CO (10-9)
line because of the three non-CH$_3$OH lines it is the one
least affected by high optical depth. Using the assumption
that emission from both of the CH$_3$OH lines' wings originate
from the same physical region of the YSO, we would expect
that when beam dilution is taken into account correctly
the line ratios of both CH$_3$OH lines divided
by the CO (10-9) would yield the same N(CH$_3$OH)/N(CO).

We decided to adopt $\alpha=1.0$ as the beam-size correction
exponent for all our cases. According to \citet{Tafalla2010}
this implies that the emission we are probing is extended
along one dimension when compared to our beam sizes.

\section{Rotational diagrams for CH$_3$OH}
\label{app:rotdia}
\begin{figure}
 \includegraphics[width=0.95\linewidth]{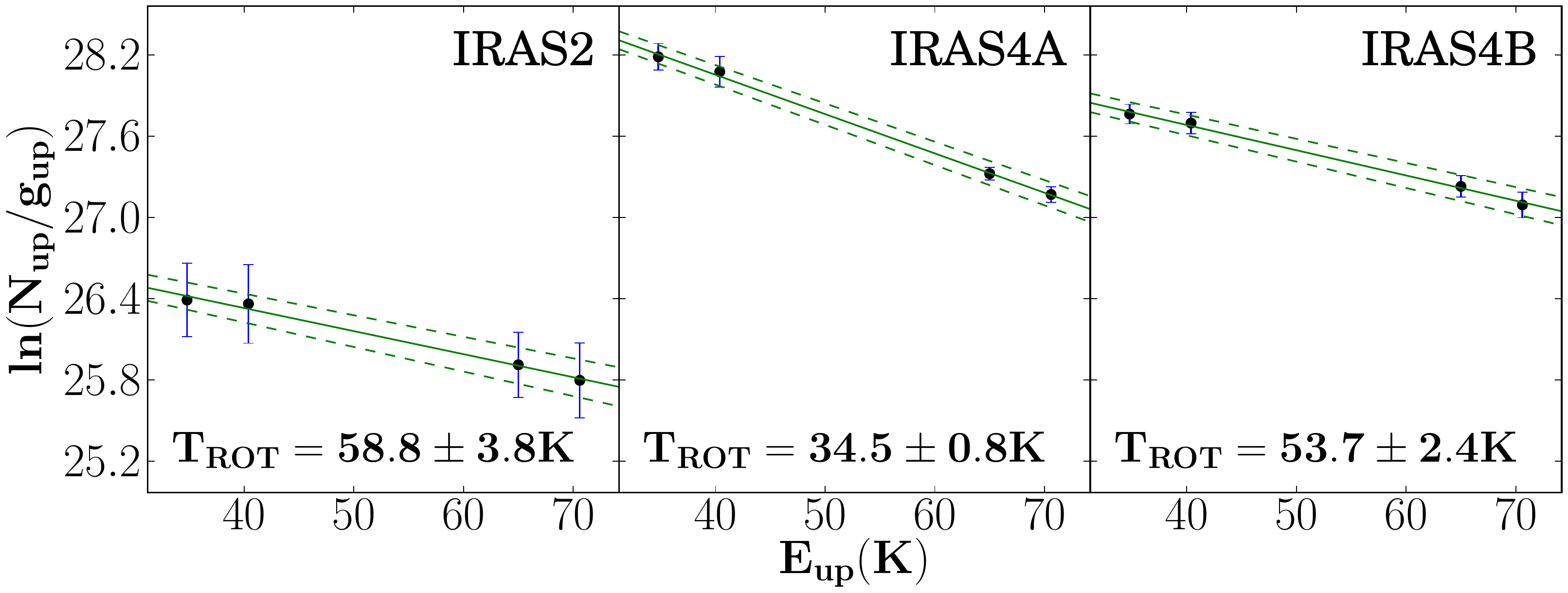}
 \caption{Rotational diagrams of CH$_3$OH for the integrated
 intensity of both wings 
 in IRAS 2A (left), IRAS 4A (middle), and IRAS 4B (right).
 The solid green line is the linear fit to the shown points and
 the dashed green lines show the
 uncertainty range of the fit as determined by the 1-sigma uncertainties
 of the fit parameters.}
 \label{fig:rotdia}
\end{figure}

We see emission from several transitions in both A-CH$_3$OH 
and A-CH$_3$OH.
Though most of the lines are very weak and/or 
blended with each other, we can use the four strongest
lines (${\rm A-CH_3OH}(5_0-4_0)$,${\rm A-CH_3OH}(7_0-7_0)$,
${\rm E-CH_3OH}(5_{-1}-4_{-1})$ and 
${\rm E-CH_3OH}(7_{-1}-7_{-1})$) to construct rotational
diagrams, which are presented in Figure \ref{fig:rotdia}
for all three objects. We calculate
$T_{rot} = 57.8\pm 4.3 K$ for IRAS 2A, 
$T_{rot} = 34.6\pm 1.3 K$ for IRAS 4A, and 
$T_{rot} = 53.7\pm 2.0 K$ for IRAS 4B.
It is also possible to construct rotational diagrams
separately for the red and blue wings of the three
objects. In doing so one ends up with
$T_{rot,red} = 91.5\pm 32.7 K$ and
$T_{rot,blue} = 46.9\pm 2.7 K$ for IRAS 2A, 
$T_{rot,red} = 23.4\pm 2.2 K$ and
$T_{rot,blue} = 40.8\pm 2.4 K$ for IRAS 4A, and 
$T_{rot,red} = 59.9\pm 7.9 K$ and
$T_{rot,blue} = 52.3\pm 4.4 K$ for IRAS 4B.

\bsp

\label{lastpage}

\end{document}